\documentclass[structabstract]{aa}  
\usepackage{longtable}
\usepackage{appendix}
\usepackage{natbib}
\usepackage{lscape}

\bibpunct{(}{)}{;}{a}{,}{,}

\usepackage{graphicx}
\usepackage{txfonts}
\begin{document}

   \title{Ionization fraction and the enhanced sulfur chemistry in Barnard 1}

   \author{A. Fuente\inst{1}
     \and 
       J. Cernicharo\inst{2}
     \and
      E. Roueff\inst{3,4}
     \and
      M. Gerin\inst{3,4}
     \and
      J. Pety\inst{5}
      \and
      N. Marcelino\inst{6}
     \and 
     R. Bachiller\inst{1}
     \and
     B. Lefloch\inst{7} 
    \and
     O. Roncero\inst{8} 
    \and
     A. Aguado\inst{9} 
   }
   \institute{Observatorio Astron\'omico Nacional (OAN,IGN), Apdo 112, E-28803 Alcal\'a de Henares (Spain)
   \email{a.fuente@oan.es}
   \and
    Instituto de Ciencia de Materiales de Madrid, ICMM-CSIC, C/ Sor Juana In\'es de la Cruz 3, 
    E-28049 Cantoblanco, Spain  
   \and
   CNRS UMR 8112, LERMA, Observatoire de Paris and \'Ecole Normale Sup\'erieure. 24 rue Lhomond, 75231 Paris Cedex 05, France
   \and
   Sorbonne Universit\'es, UPMC Univ. Paris 06, UMR8112, LERMA, F-75005 Paris, France 
   \and
   Institut de Radioastronomie Millim\'etrique, 300 Rue de la Piscine, F-38406 Saint Martin d'H\'eres, France
   \and
  INAF, Osservatorio di Radioastronomia, via P. Gobetti 101, I-40129, Bologna, Italy
   \and
   Institut de Plan\'etologie et d'Astrophysique de Grenoble (IPAG) UMR 5274, Université UJF-Grenoble 1/CNRS-INSU, F-38041 Grenoble, France   
   \and
    Instituto de F\'{\i}sica Fundamental (IFF-CSIC), C.S.I.C., Serrano 123, E-28006 Madrid, Spain
    \and
    Facultad de Ciencias, Unidad Asociada de Qu\'{\i}mica-F\'{\i}sica Aplicada CSIC-UAM, Universidad Aut\'onoma de Madrid, E-28049 Madrid, Spain
}


 
  \abstract
  %
  {Barnard B1b has been revealed as one of the most
interesting globules from the chemical and dynamical point of view. It presents a rich molecular chemistry characterized by
large abundances of deuterated and complex molecules. Furthermore, this globule hosts an extremely young Class 0 object and one candidate for  
the first hydrostatic core (FHSC) proving the youth of this star-forming region.}
  {Our aim is to determine the cosmic ray ionization rate, $\zeta_{H_2}$, and the 
depletion factors in this extremely young star-forming region.
These parameters determine the dynamical evolution of the core.}
    {We carried out a spectral survey toward Barnard 1b
as part of the IRAM large program $``$IRAM Chemical survey of sun-like star-forming regions$"$ (ASAI) 
using the IRAM 30-m telescope at Pico Veleta (Spain). 
This provided a very complete inventory
of neutral and ionic C-, N-, and S- bearing species with, from our knowledge,
  the first secure detections of the deuterated ions DCS$^+$ and DOCO$^+$. 
We use a state-of-the-art pseudo-time-dependent gas-phase chemical model that 
includes the ortho and para forms of
H$_2$, H$_2^+$, D$_2^+$, H$_3^+$, H$_2$D$^+$, D$_2$H$^+$, D$_2$, and D$_3^+$ 
to determine the local value of the cosmic ray ionization rate and the depletion 
factors.}
  {Our model assumes n(H$_2$)=10$^5$~cm$^{-3}$ and T$_k$=12~K, as derived from 
our previous works.
  The observational data are well fitted with $\zeta_{H_2}$ between 
3$\times$10$^{-17}$~s$^{-1}$ and 10$^{-16}$~s$^{-1}$
and the elemental abundances 
  O/H=3~10$^{-5}$, N/H=6.4$-$8~10$^{-5}$, C/H=1.7~10$^{-5}$, and S/H between 
6.0~10$^{-7}$ and 1.0~10$^{-6}$. The large number of 
  neutral/protonated species detected allows us to derive the elemental 
abundances and cosmic ray ionization rate simultaneously. 
Elemental depletions are estimated to be $\sim$10 for C and O, $\sim$1 for N,  
and $\sim$25 for S. 
}
   {Barnard B1b presents similar depletions of C and O as those measured in 
prestellar cores. 
   The depletion of sulfur is higher than that of C and O, but not as extreme as 
in cold cores.
   In fact, it is similar to the values found in some bipolar outflows, hot 
cores, and 
   photon-dominated regions. Several scenarios are discussed to account for 
these peculiar 
   abundances. We propose that it is the consequence of the initial conditions 
(important outflows and enhanced UV fields in the surroundings) and a rapid collapse ($\sim$0.1~Myr) that allows most S- and N-bearing species to remain 
   in the gas phase to great optical depths. The interaction of the 
compact outflow associated with B1b-S with the surrounding material could enhance the abundances of S-bearing 
molecules, as well.
}


   \maketitle


\section{Introduction}
Cosmic rays (hereafter CRs) play a leading role in the chemistry, heating, and 
dynamics 
of the interstellar medium (ISM). By 
ionizing their main
component,\ molecular hydrogen, CRs interact with dense molecular clouds,  and this process initiates a series of 
ion-molecule reactions
that form compounds of increasing complexity to eventually  
build up the rich chemistry observed in dark clouds (see, e.g.,
\citealp{Wakelam10}).
Also, CRs represent an important heating source for molecular clouds. Collisions 
with interstellar molecules and atoms convert about half of the energy of 
primary and secondary 
electrons yielded by the ionization process into heat 
\citep{Glassgold73,Glassgold12}.
Finally, the ionization fraction controls the coupling of magnetic fields with 
the gas, 
driving the dissipation of turbulence and angular momentum transfer, thus 
playing a crucial role 
in protostellar collapse and the dynamics of accretion disks (see 
\citealp{Padovani13} and references 
therein).

\begin{figure}
\hspace*{-0.0cm}
\includegraphics[scale=0.59,angle=0]{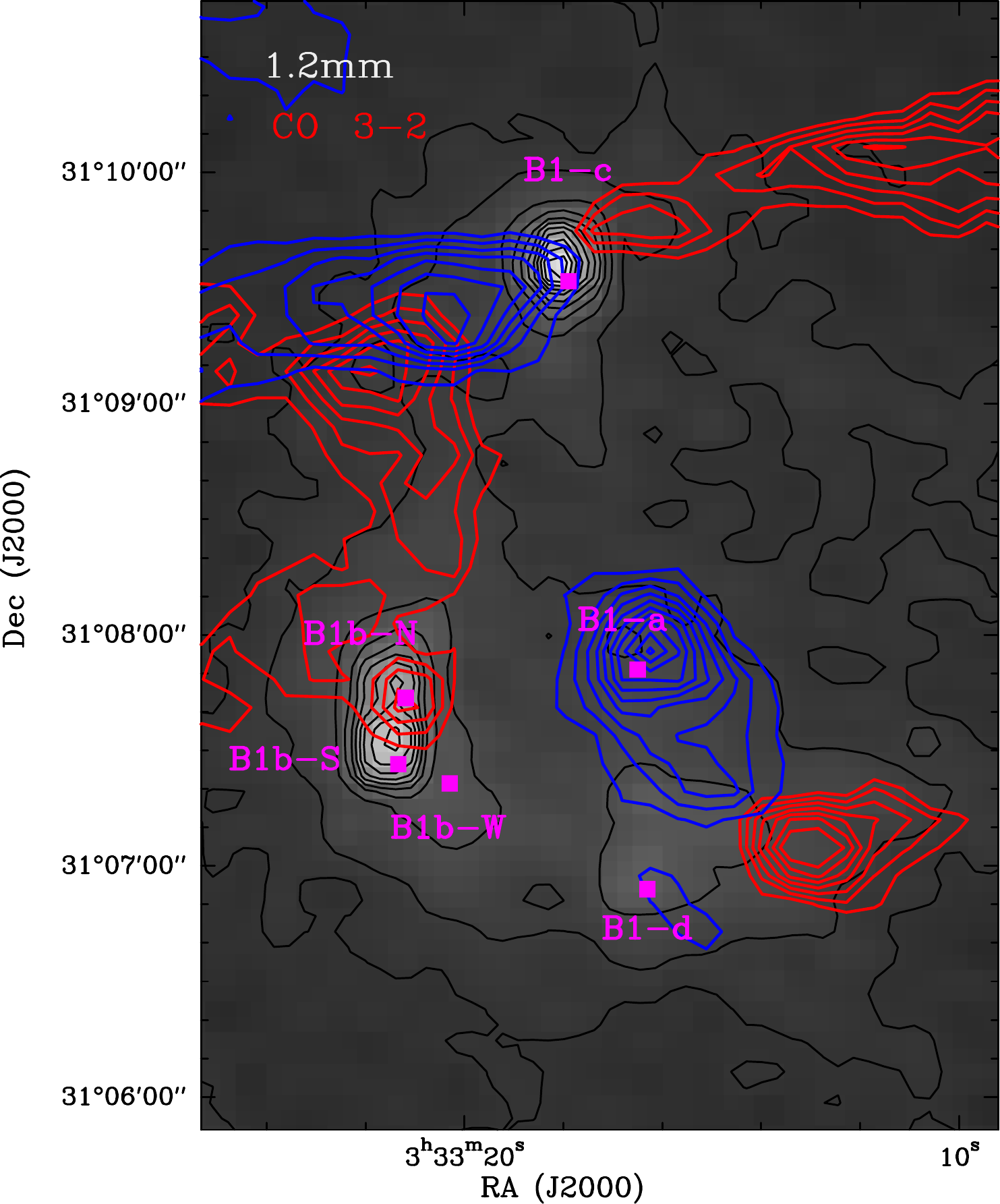}
   \caption{\label{fig1}
Continuum map at 1.2 mm observed at the
IRAM telescope  with the  MAMBO  instrument \citep{Daniel13}. The  contours  are 
20mJy/beam, 
40 mJy/beam to 400 mJy/beam in steps of by 40 mJy/beam. Contours show the high 
velocity 
blue (2.5 to 4.5 km s$^{-1}$) and red (8.5 to 11.5 km s$^{-1}$) emission of 
the CO 3$\rightarrow$2 line
as observed with the JCMT (project S12AC01).
}
\end{figure}

The cosmic ray ionization rate $\zeta _{H_2}$ is the number of hydrogen molecule 
ionization 
per second produced by CRs. In absence of other ionization agents (X-rays, UV 
photons, J-type shocks), 
the ionization fraction is proportional to $\sqrt{\zeta _{H_2}}$, which becomes 
the key parameter in
the molecular cloud evolution \citep{Kee89,Caselli02}.
During the last 50 years, values of $\zeta _{H_2}$ ranging from a 
few 10$^{-18}$ s$^{-1}$ to a few 10$^{-16}$ s$^{-1}$ have 
been observationally determined in diffuse and dense interstellar clouds 
(Galli \& Padovani 2015, and references therein). 
The cosmic ray ionization rate has usually been estimated from the
[HCO$^+$]/[CO] and/or [DCO$^+$]/[HCO$^+$] abundance ratios 
using simple analytical expressions (e.g., 
\citealp{Guelin77,Wootten79,Dalgarno84}). 
In recent studies, these estimates have been improved using
comprehensive chemistry models \citep{Caselli99,Bergin99}. 
These models show that the gas ionization fraction is also dependent on the 
depletion factors
that should be known to accurately estimate $\zeta _{H_2}$ \citep{Caselli99}. 

Chemical modeling is required to constrain the elemental
abundances and the cosmic ray ionization rate.
We carried out a spectral survey toward Barnard 1b as part of the IRAM large 
program $``$IRAM Chemical survey of sun-like 
star-forming regions$"$ (ASAI) (PIs: Rafael Bachiller, Bertrand Lefloch) using the IRAM 30-m telescope 
at Pico Veleta (Spain), which provided a very complete inventory of
ionic species including C-, N-, and S-bearing compounds.
The large number of neutral/protonated species allows us
to derive the elemental abundances in the gas phase and cosmic ray ionization 
rate simultaneously, avoiding 
the uncertainty due to unknown depletion factors in the latter.

\begin{figure*}
\hspace*{-0.0cm}
\includegraphics[scale=0.75,angle=0]{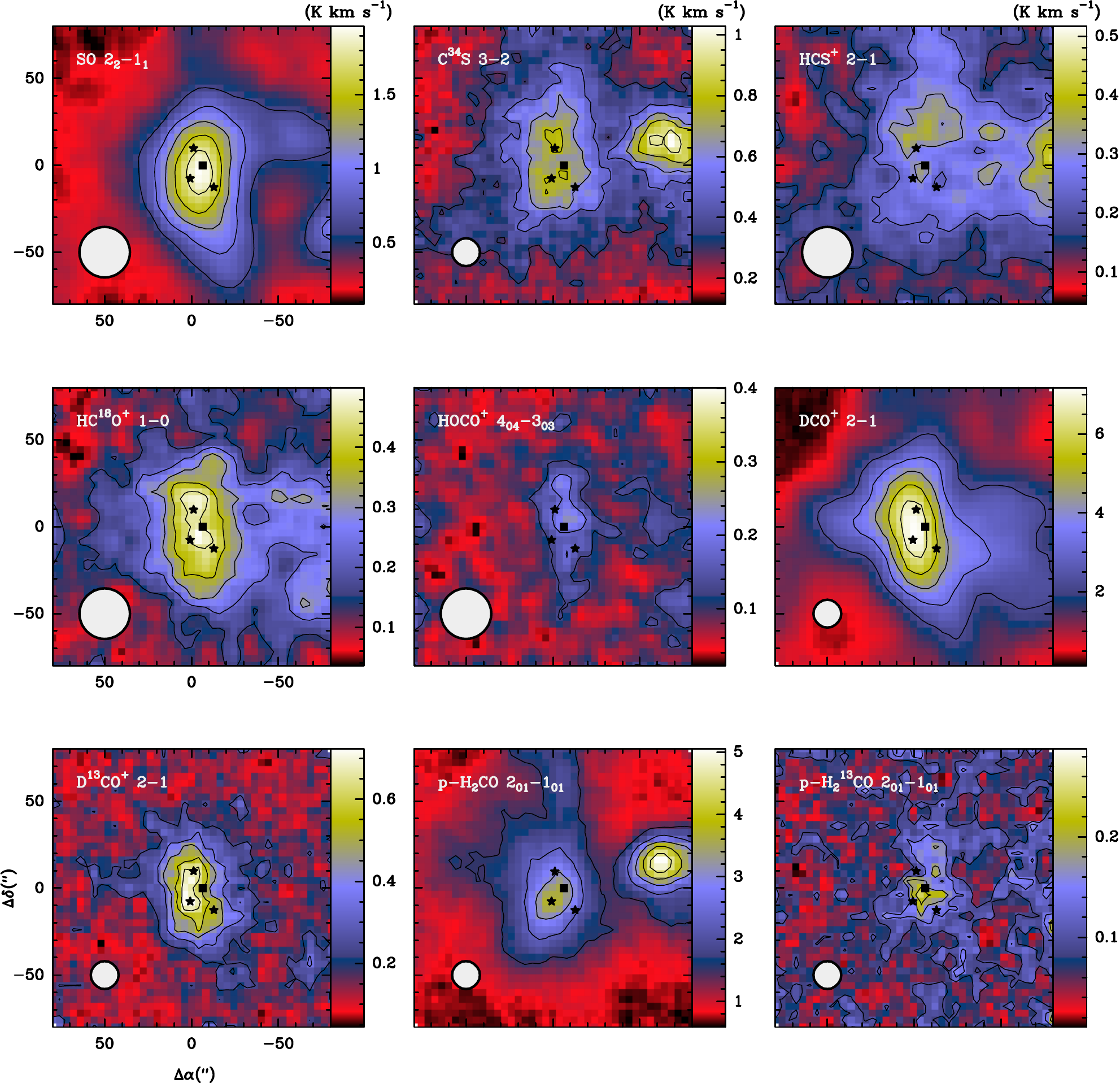}
   \caption{\label{fig2}
Integrated intensity maps of the SO 2$_2$$\rightarrow$1$_1$, 
C$^{34}$S 3$\rightarrow$2, 
HCS$^+$ 2$\rightarrow$1, HC$^{18}$O$^+$ 1$\rightarrow$0,
HOCO$^+$ 4$_{04}$$\rightarrow$3$_{03}$, DCO$^+$ 2$\rightarrow$1, 
D$^{13}$CO$^+$ 2$\rightarrow$1, p-H$_2$CO 2$_{02}$$\rightarrow$2$_{01}$, and
p-H$_2$$^{13}$CO 2$_{02}$$\rightarrow$2$_{01}$ lines. The HPBW of the 30-m 
telescope at each line frequency is drawn in the 
bottom-left corner of the panel.
The stars indicate the positions of B1b-N, B1b-S and B1b-W. The black square 
shows the position targeted in the
unbiased spectral survey. 
}
 \end{figure*}

Barnard 1 (B1) belongs to the Perseus molecular cloud complex (D=235~pc). 
The Barnard 1 dark cloud itself is subdivided
into several dense cores at different evolutionary stages of star formation. 
While B1-a and B1-c are known
to host class 0 sources associated with high velocity outflows 
\citep{Hatchell05,Hatchell07}, the B1-b core appears to be more
quiet, although its western edge is interacting with an outflow that is possibly arising 
from sources B1-a or B1-d,
which are both located 1$\arcmin$ SW of B1-b (see a sketch in Fig.~\ref{fig1}). 

The B1b region was mapped in
many molecular tracers, for example, HCO$^+$, CS, NH$_3$, $^{13}$CO 
\citep{Bachiller84,Bachiller90}, 
N$_2$H$^+$ \citep{Huang13}, H$^{13}$CO$^+$ \citep{Hirano99}, 
or CH$_3$OH \citep{Hiramatsu10,Oberg10}. 
Interferometric observations of Barnard 1b revealed that this compact core hosts 
two young stellar objects (YSOs),
B1b-N and B1b-S \citep{Huang13}, and a third more evolved source, hereafter 
B1b-W, 
with deep absorption from ices \citep{Jorgensen06}.
The two first sources are deeply embedded in the surrounding protostellar 
envelope, which
seems essentially unaffected by the embedded sources as shown by the large 
column density, 
N(H$_2$)$\sim$7.6$\times$10$^{22}$~cm$^{-2}$ \citep{Daniel13}, and cold kinetic 
temperature, T$_K$ = 12~K \citep{Lis10}.
Based on Herschel fluxes and subsequent SED modeling, \citet{Pezzuto12} 
concluded that B1b-N and 
B1b-S were younger than Class 0 sources and proposed 
these as first hydrostatic core (FHSC) candidates. Using NOEMA, 
\citet{Gerin15} obtained high resolution maps
of the protostellar ouflows and their envelope, confirming the young ages
of B1b-N and B1b-S ($\sim$1000 yr and $\sim$3000 yr) and the high density
of the surrounding core ($\sim$ a few 10$^5$ cm$^{-3}$ for the outflowing gas). 

From the chemical point of view, B1b is characterized by a rich chemistry. 
Indeed, many molecules were observed for the first time in this object, like
HCNO \citep{Marcelino09} or CH$_3$O \citep{Cernicharo12}. Additionally, B1b 
shows a high degree of deuterium fractionation
and has been associated with first detections of multiply
deuterated molecules, such as ND$_3$ \citep{Lis02} or
D$_2$CS \citep{Marcelino05}, which is consistent with the expected chemistry in a
dense and cold core.

\begin{table}
\caption{Observing and telescope parameters.}

\label{tab_observations}
\begin{tabular}{ccccc} \hline \hline
 \multicolumn{1}{c}{Freq (GHz)}  &      \multicolumn{1}{c}{$\Delta$v 
(km~s$^{-1}$)} &  \multicolumn{1}{c}{HPBW$^1$} &  \multicolumn{1}{c}{$B_{eff}$} & 
 \multicolumn{1}{c}{rms(T$_a$) (mK)} \\ \hline
82.5$-$117.6                     &       0.1255                             &   
29$\arcsec$$-$21$\arcsec$   & 0.81$-$0.78  &  3$-$5$^2$ \\
130.0$-$172.7                    &       0.3424                             &   
19$\arcsec$$-$14$\arcsec$   & 0.74 &  6$-$10$^3$ \\
200.6$-$275.9                    &       0.2839                             &   
12$\arcsec$$-$9$\arcsec$    & 0.63$-$0.49 & 5$-$10 \\ \hline
\end{tabular}

\noindent
$^1${\scriptsize HPBW(arcsec)$\approx$2460/Freq(GHz);} $^2${\scriptsize rms$\sim$240~mK for $\nu$$>$114~GHz,;}
$^3${\scriptsize rms$\sim$500~mK for $\nu$$>$169~GHz. }
\end{table}

\begin{figure*}
\hspace*{-0.0cm}
\includegraphics[scale=0.75,angle=0]{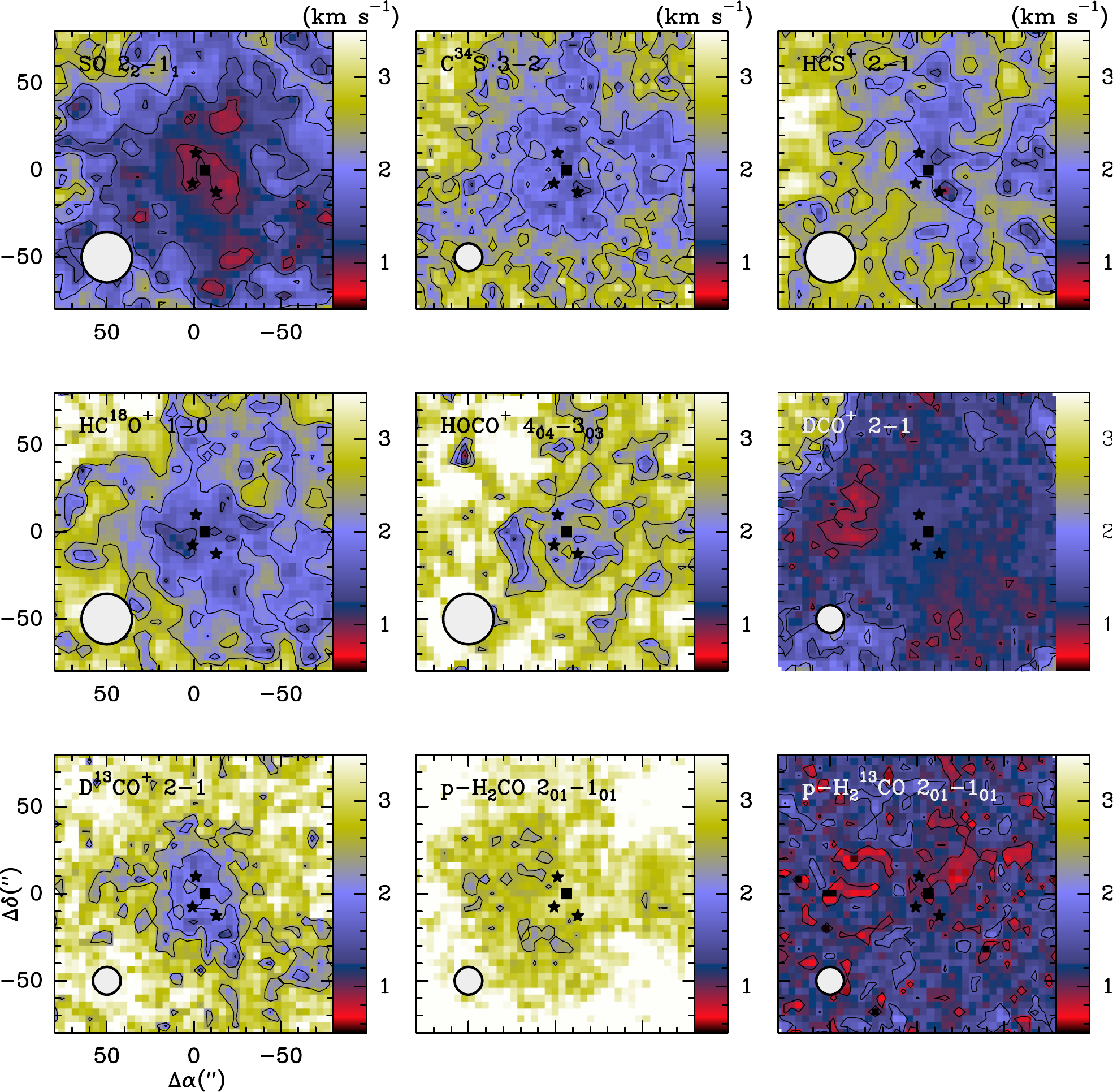}
   \caption{\label{fig3}
Maps of the linewidths of the SO 2$_2$$\rightarrow$1$_1$, 
C$^{34}$S 3$\rightarrow$2, HCS$^+$ 2$\rightarrow$1, HC$^{18}$O$^+$ 
1$\rightarrow$0,
HOCO$^+$ 4$_{04}$$\rightarrow$3$_{03}$, DCO$^+$ 2$\rightarrow$1, 
D$^{13}$CO$^+$ 2$\rightarrow$1, p-H$_2$CO 2$_{02}$$\rightarrow$2$_{01}$,
p-H$_2$$^{13}$CO 2$_{02}$$\rightarrow$2$_{01}$ lines. Symbols are the same as in 
Fig.~\ref{fig2}.
}
 \end{figure*}

\begin{figure*}
\hspace*{-0.0cm}
\includegraphics[scale=0.75,angle=0]{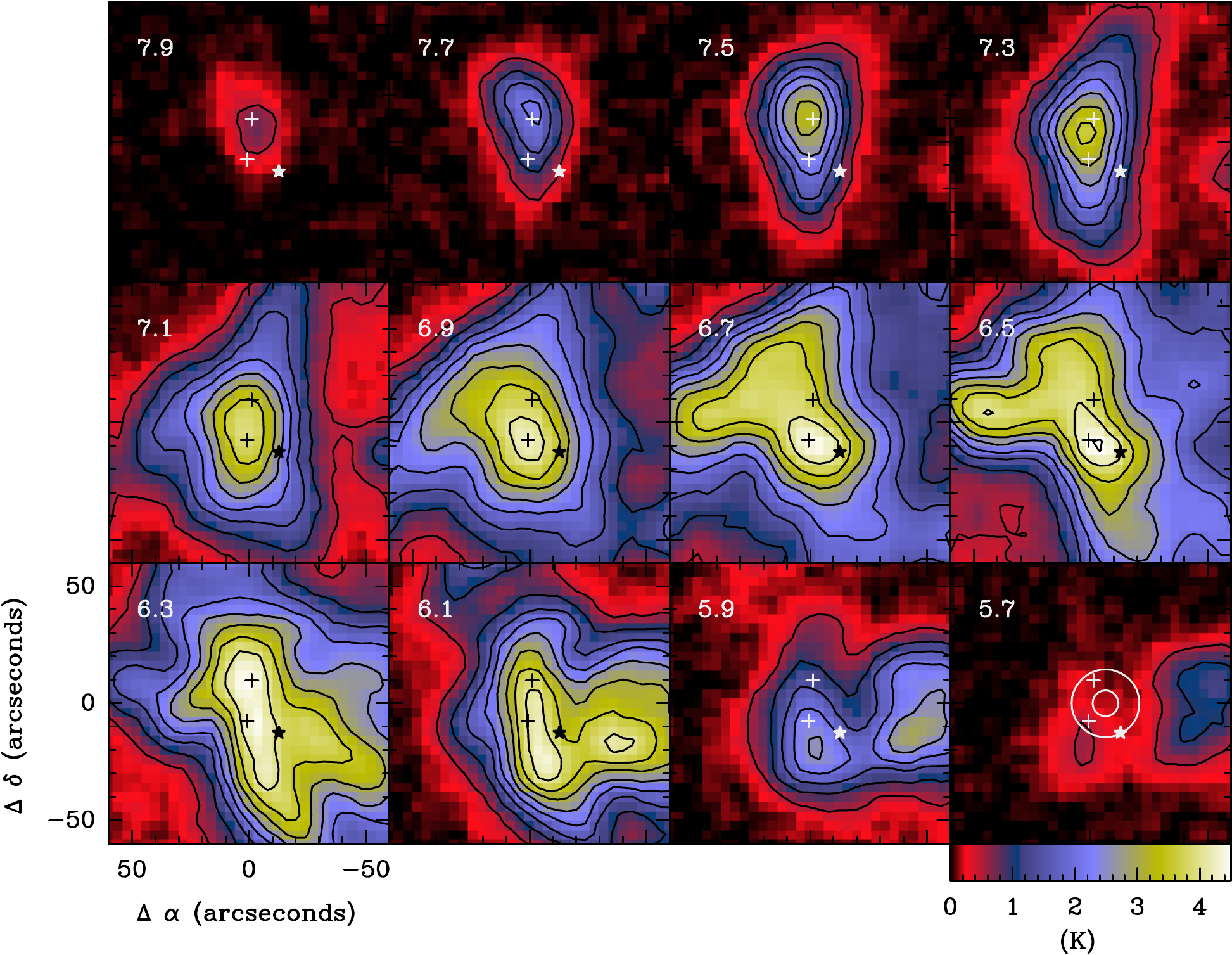}
   \caption{\label{fig4}
Spectral maps of the  DCO$^+$ 2$\rightarrow$1 line as observed 
with the IRAM 30-m telescope. The central
velocity is indicated in the top-left corner. Contours are 0.5~K to 4.5~K by 
steps of 0.5~K. The crosses indicate the positions
of the protostellar cores B1b-N and B1b-S. The star denotes the position of B1b-W. 
In the last panel we plot the beam at 144.277~GHz.
}
\end{figure*}

\begin{figure*}
\hspace*{-0.0cm}
\includegraphics[scale=0.75,angle=0]{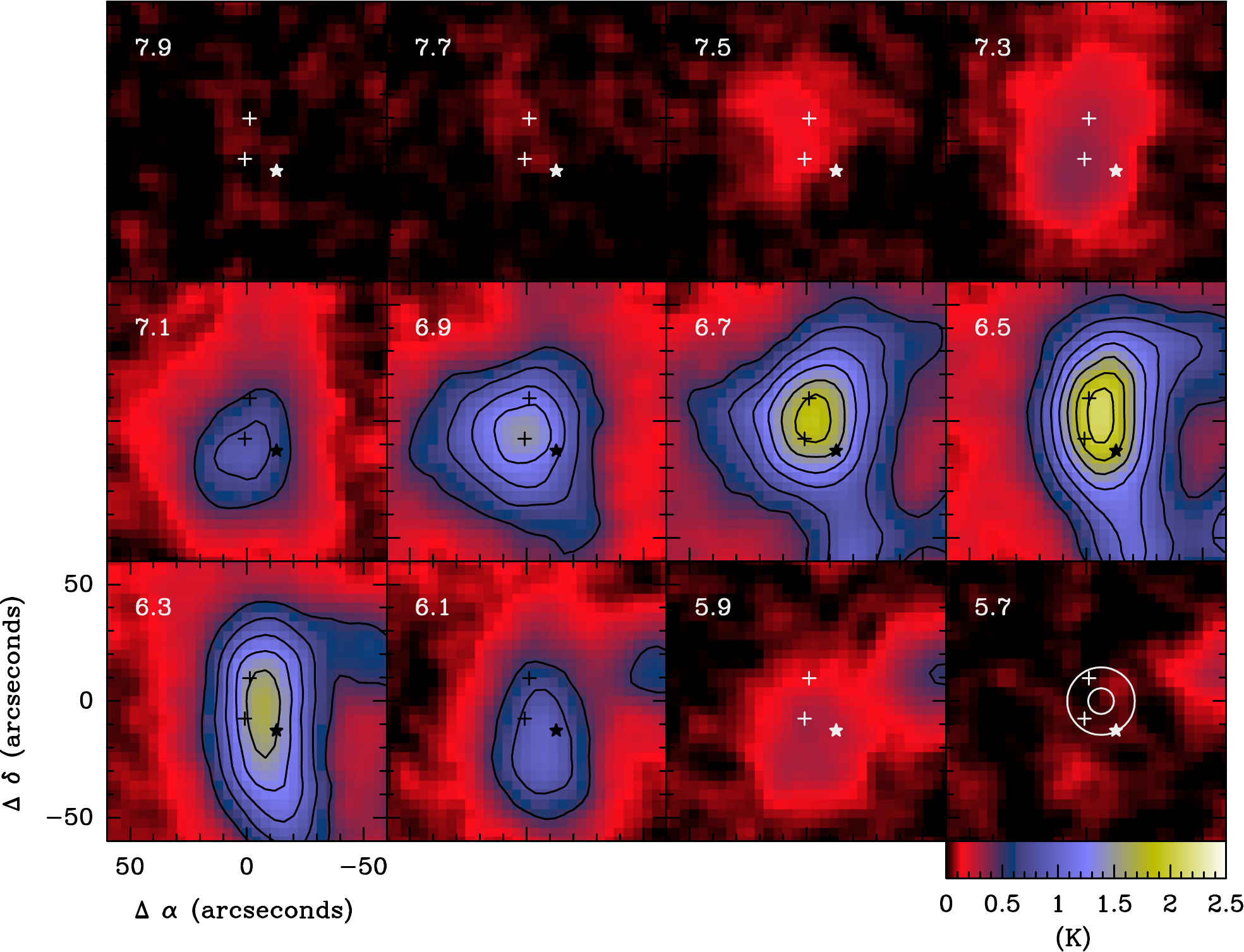}
   \caption{\label{fig5}
Spectral maps of the SO 2$_2$$\rightarrow$1$_1$ line as observed 
with the IRAM 30-m telescope. The central
velocity is indicated in the top-left corner. Contours are 0.5~K to 4.5~K by 
steps of 0.25~K. The crosses indicate the positions
of the protostellar cores B1b-N and B1b-S.The star denotes the position of B1b-W. 
In the last panel we plot the beam at 86.094~GHz.
}
\end{figure*}

\section{Observations}
We carried out a spectral survey toward Barnard B1b
($\alpha_{\rm J2000}$=03$\mathrm{h}$\,33\,$\mathrm{m}$ 20.8\,$\mathrm{s}$,\, 
$\delta_{\rm J2000}$=31$^{\circ}$ 07' 34''.0) 
using the IRAM 30-m telescope at Pico Veleta (Spain). This position is in 
between
the two FHSC candidates B1b-N and B1b-S. The two protostellar cores lie inside 
the HPBW of the 30-m telescope
at 3 mm, but they are missed at 1 mm wavelengths (see Fig.~\ref{fig2}). 

The 3 mm observations were performed between January and May 2012
(see \citealp{Cernicharo12}), which used the narrow mode of the FTS
allowing a higher spectral resolution of 50 kHz. The observing mode
in this case was frequency switching with a frequency throw of 7.14 MHz,
which removes standing waves between the secondary and the receivers.
Several observing periods were scheduled on July and February 2013 and  
January and March 2014 to cover the 1 and 2 mm bands shown in 
Table~\ref{tab_observations}. 
The Eight MIxer Receivers (EMIR) and the Fast Fourier Transform Spectrometers 
(FTS) with a spectral resolution of 200\,kHz
were used for these observations.
The observing procedure was wobbler switching with a throw of 200$\arcsec$ to 
ensure flat baselines. In
Table~\ref{tab_observations} we show the half power beam width (HPBW), spectral 
resolution, and sensitivity 
achieved in each observed frequency band. 
The whole dataset will be published elsewhere but
this paper is centered on the ionic species and sulfur compounds. 

We complemented the unbiased spectral survey with 
175$\arcsec$$\times$175$\arcsec$ maps of 
the SO 2$_2$$\rightarrow$1$_1$, C$^{34}$S 3$\rightarrow$2, HCS$^+$ 
2$\rightarrow$1, HC$^{18}$O$^+$ 1$\rightarrow$0,
HOCO$^+$ 4$_{04}$$\rightarrow$3$_{03}$, DCO$^+$ 2$\rightarrow$1, 
D$^{13}$CO$^+$ 2$\rightarrow$1, p-H$_2$CO 2$_{02}$$\rightarrow$2$_{01}$,
p-H$_2$$^{13}$CO 2$_{02}$$\rightarrow$2$_{01}$ lines (see Fig.~\ref{fig2}).  
These maps were simultaneously observed at 3 and
2~mm with a combination of the EMIR receivers and the Fourier
transform spectrometers, which yields a bandwidth of 1.8~GHz
per sideband per polarization at a channel spacing of 49~kHz.
The tuned frequencies were 85.55 and 144.90~GHz. We used the
on-the-fly scanning strategy with a dump time of 0.5 seconds
and a scanning speed of 10.9$\arcsec$/s to ensure a sampling of at least
three dumps per beam at the 17.8$\arcsec$ resolution of the 2 mm lines. The
175$\arcsec$$\times$175$\arcsec$ region was covered using successive orthogonal 
scans
along the RA and DEC axes. The separation between two successive rasters 
was 6.5$\arcsec$ ($\sim$ $\lambda$/2D) to ensure Nyquist sampling
at the highest observed frequency, i.e., 145~GHz. A common reference position 
located at offsets (200$\arcsec$, 150$\arcsec$) was observed for
10 seconds every 77.5 seconds. The typical IRAM 30 m position
accuracy is 3$\arcsec$.

The data reduction and line identification were carried out with the 
package \texttt{GILDAS}\footnote{See \texttt{http://www.iram.fr/IRAMFR/GILDAS} 
for
  more information about the GILDAS
  softwares~\citep{pety05}.}\texttt{/CLASS} software. 

\begin{figure}
\hspace*{-0.0cm}
\includegraphics[scale=0.59,angle=0]{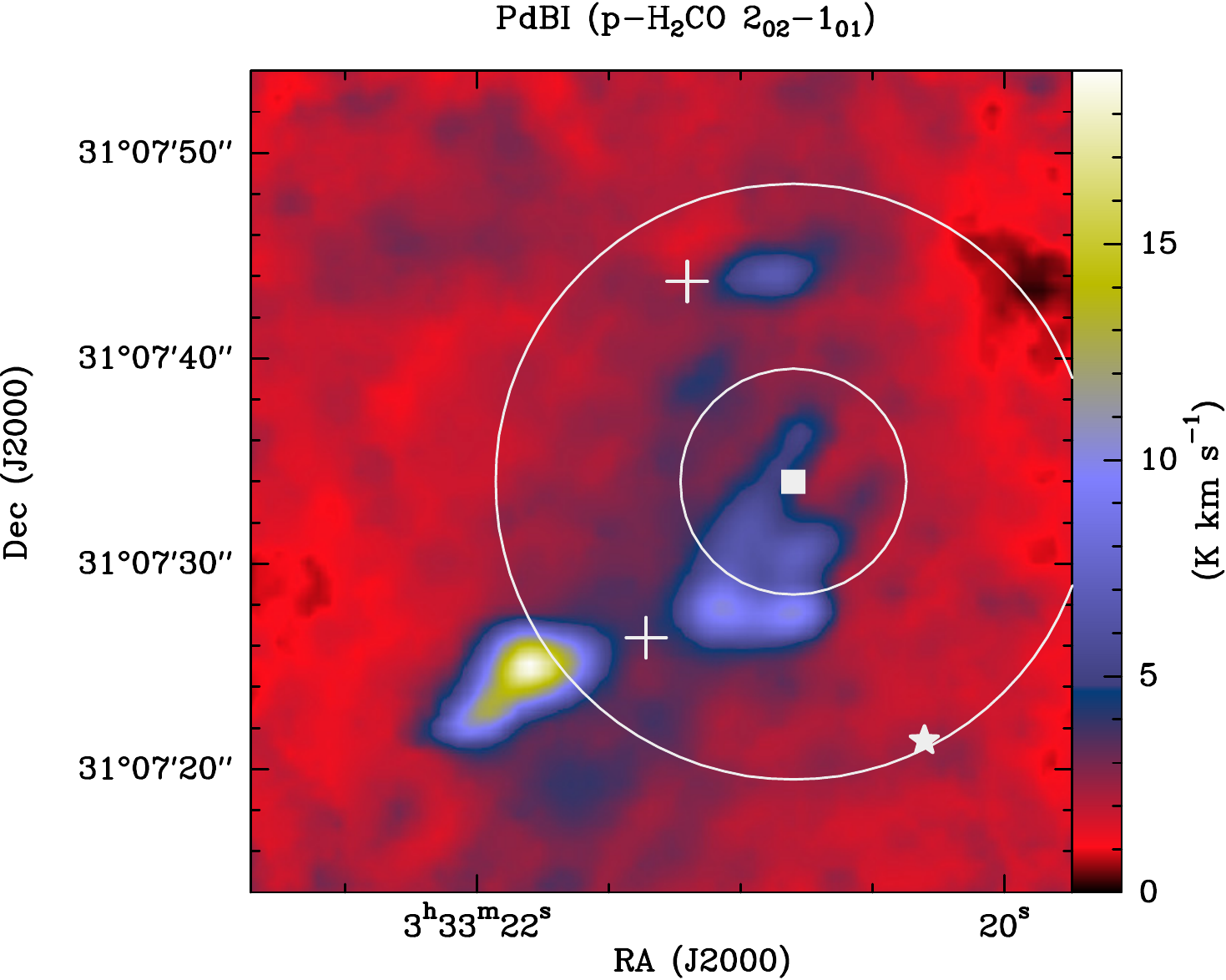}
   \caption{\label{fig6}
Interferometric integrated intensity map of the 
outflows associated with B1b-N and B1b-S in the H$_2$CO 2$_{0,2}$$-$1$_{0,1}$ 
line \citep{Gerin15}. The crosses indicate the positions
of the protostellar cores, B1b-N and B1b-S, and the star, of B1b-W. The position 
targeted by our spectral survey is
drawn with a white square and the beams of the 30-m telescope at 3nmm and 1mm are 
indicated by white circles.
}
\end{figure}

\section{Methodology}
Our goal is to use the database provided by the IRAM large program ASAI together with previous millimeter
data from our group \citep{Marcelino05, Marcelino09, Marcelino10, Cernicharo12,
Cernicharo13, Daniel13, Cernicharo14, Gerin15}  and complementary mapping to 
accurately determine the fundamental chemical parameters.  

Within our spectral survey we detected the following ions:
HCO$^+$, H$^{13}$CO$^+$, HC$^{18}$O$^+$, HC$^{17}$O$^+$, DCO$^+$, 
D$^{13}$CO$^+$,
HOCO$^+$, DOCO$^+$, HCS$^+$, SO$^+$, HC$^{34}$S$^+$, DCS$^+$,
HCNH$^+$, H$_2$COH$^+$, HC$_3$NH$^+$,
NNH$^+$, NND$^+$, N$^{15}$NH$^+$, and $^{15}$NNH$^+$. 
This is probably the first secure detection of the deuterated ions DCS$^+$ and 
DOCO$^+$.
A tentative detection of DOCO$^+$ was reported by \citet{Fech13} toward the 
massive protostar
CIGX-N63. \citet{Daniel13} modeled the emission of NNH$^+$, NND$^+$, 
N$^{15}$NH$^+$, 
and $^{15}$NNH$^+$ ions in this core. 
We use their results to compare with our chemical models without repeating the 
analysis. 
Our final list of detections is the most
complete inventory of molecular ions observed in this source and, 
therefore, an unprecedented opportunity to improve our knowledge of the 
ionization
fraction.
The main destruction mechanism of the observed protonated ions is the rapid 
recombination with $e^-$ to
give back the neutral parent species. For this reason, the [XH$^+$]/[X] ratios 
are very
sensitive to the gas ionization fraction and are commonly used to estimate it. 
 We also include the lines of 
the parent molecules in our analysis to compare with chemical models (see Table.~A.1). It is interesting to note 
that we have not detected CO$^+$ and HOC$^+$. These
ions react with molecular hydrogen and only reach significant abundances in the 
external
layers of photon-dominated regions that are not fully molecular (see, e.g., 
\citealp{Fuente03,Rizzo03}).
The non-detection of these ions mainly supports the interpretation that we are
probing a dense molecular core in which the ionization 
is dominated by cosmic rays.

The second key ingredient in gas-phase chemical models is the depletion factors.
The sensitivity of the chemistry to the elemental abundances in the gas phase was
discussed by \citet{Graedel82}, who introduced the concept of $``$low-metal$"$ 
elemental
abundances, in which the abundances of metals, silicon, and sulfur are depleted 
by a
factor of 100 with respect to the values typically used until then (those 
observed in the
diffuse cloud $\zeta$ Oph). This choice of elemental
abundances provides a reasonable agreement with the observed molecular 
abundances in dark 
clouds (see review by \citet{Agundez13}). In recent decades, 
there has been increasing evidence
that these abundances are decreasing with density because  
molecules are frozen onto the grain surfaces and reach the lowest values in 
dense prestellar 
cores. This phenomenon is quantified by the so-called depletion factor, that is, 
the ratio between 
the total (ice plus gas) abundance and that observed in the gas phase. The depletion 
is
defined for a specific molecule or for the total amount of atoms of a given 
element.
The depletion factor of each species depends on the 
local physical conditions, time evolution, and binding energy. 
For a given molecule, the depletion factor is evaluated by measuring the gas-phase abundance 
as a function of the gas density within the core. \citet{Tafalla06} have shown 
that 
CO presents a strong variation while the effect is moderate
for HCN and almost null for N$_2$H$^+$.

Determining the depletion factor of a given element is more complex, since it
depends on the gas chemical composition and, eventually, on the binding energy 
of 
the compound(s) that are the most abundant. To derive the elemental
gas abundances of C, O, N, and S, we need to include the main
reservoirs
of these elements in our analysis. The CO abundance is four orders of magnitude larger than that 
of any other 
molecule except H$_2$. Therefore, we can assume that in a dense core, such as 
Barnard 1b,
all the carbon is locked in CO. 
The main reservoirs of nitrogen are atomic nitrogen (N) and molecular nitrogen 
(N$_2$), which are not observable. 
The nitrogen abundance can be derived by applying a chemical model and fitting 
the observed abundances
of nitriles (HCN, HNC, and CN) and N$_2$H$^+$. 
The most abundant oxygenated molecules,
O$_2$, H$_2$O, and OH, cannot be observed in the millimeter domain 
and the oxygen depletion factor has to be derived indirectly. 
In the case of sulfur, the abundances of S-bearing
species are very sensitive to the C/O gas-phase ratio, and hence
the oxygen depletion factor. 
A complete chemical modeling is needed to have a reasonable estimate 
of the O and S depletion factors.  

\section{Results}

Figures~\ref{figB1} to ~\ref{figB9} show the spectra of the neutral and protonated 
species considered in
this work. In all cases, the protonated and neutral species present similar line 
profiles suggesting that 
both come from the same region. All the lines show two narrow velocity 
components, at $\sim$6.5~km~s$^{-1}$ and $\sim$7.0~km~s$^{-1}$, respectively. 
We use the 30 m maps to explore the origin of the different species and the two 
velocity components. 
The integrated intensity maps of the SO 2$_2$$\rightarrow$1$_1$, C$^{34}$S 
3$\rightarrow$2, 
HCS$^+$ 2$\rightarrow$1, HC$^{18}$O$^+$ 1$\rightarrow$0,
HOCO$^+$ 4$_{04}$$\rightarrow$3$_{03}$, DCO$^+$ 2$\rightarrow$1, 
D$^{13}$CO$^+$ 2$\rightarrow$1, p-H$_2$CO 2$_{02}$$\rightarrow$2$_{01}$, and
p-H$_2$$^{13}$CO 2$_{02}$$\rightarrow$2$_{01}$ lines are shown in 
Fig.~\ref{fig2}.  
The emission of the HC$^{18}$O$^+$, HOCO$^+$, C$^{34}$S, DCO$^+$, D$^{13}$CO$^+$, 
and 
SO lines is concentrated in the elongated filament that contains B1b-N and B1b-S
(see  Fig.~\ref{fig2}). 
The emission of the N$_2$H$^+$ 1$\rightarrow$0 line, as reported by 
\citet{Daniel13},
is also concentrated in this dense filament.
In the case of HOCO$^+$, the elongated emission seems to be slightly shifted to 
the west although the map is too noisy to conclude. In all these cases, the line 
intensities are similar toward the N and S protostars. 
On the contrary, the emission of the o-NH$_2$D 1$_{11}s$$\rightarrow$0$_{01}a$ 
line peaks toward B1b-N \citep{Daniel13}. 
Another exception is H$_2$CO whose emission
presents an intense peak toward B1b-S. 
The emission of the HCS$^+$ 2$\rightarrow$1 line presents a flat spatial 
distribution suggesting that this emission
is coming from the cloud envelope. 

Linewidths can also provide information about the origin of the molecular 
emission.
On basis of low angular resolution observations, \citet{Bachiller90} proposed a 
correlation between the linewidths
and the depth into the B1 cloud with the narrowest lines ($\sim$1 km~s$^{-1}$) 
coming from the highest extinction regions. 
Similar results were obtained by \citet{Daniel13} on basis of the observation 
of HCN, HNC, N$_2$H$^+$, NH$_3$, and their isotopologues.
The line profiles were well reproduced by a collapsing dense core (infall 
velocity = 0.1 km~s$^{-1}$) with a turbulent velocity of 
0.4~km~s$^{-1}$ surrounded by a static lower density envelope with a turbulent 
velocity of 1.5~km~s$^{-1}$. 
In Fig.~\ref{fig3}, we show  the spatial distribution of the linewidths for the 
set of maps shown in Fig.~\ref{fig2}. In agreement with previous works, 
we find a clear decrease of the linewidth with the distance from the core center 
in the SO 2$_2$$\rightarrow$1$_1$ line.
Linewidths are $\sim$2~km~s$^{-1}$ in the most external layer and decrease 
to $\sim$0.7~km~s$^{-1}$ toward the core center. However, the linewidths of the
HCS$^+$  2$\rightarrow$1 lines are $\sim$2~km~s$^{-1}$ all over the region 
without a clear pattern. This supports the interpretation of
the HCS$^+$ emission coming from an external layer of the cloud. The only 
position where the linewidths of the HCS$^+$ line becomes $<$1~km~s$^{-1}$
is toward the IR star B1b-W. For HC$^{18}$O$^+$,
DCO$^+$, and D$^{13}$CO$^+$, we do see a trend, with the linewidths decreasing 
with the decreasing radius from the core center. 
However, their linewidths are always $>$1~km~s$^{-1}$ suggesting that the 
emission is not dominated by the innermost layers of 
the envelope. The linewidth of the p-H$_2$CO line is very constant over the 
whole core. 
This line is more likely to be optically thick and its emission is 
only tracing the outer layers of the core.

Figures~\ref{fig4} and \ref{fig5} show the spectral maps of the DCO$^+$ 
2$\rightarrow$1 and 
SO 2$_2$$\rightarrow$1$_1$ lines, respectively. The emission of DCO$^+$ 
extends from $\sim$5.5~km~s$^{-1}$ to  $\sim$8.0~km~s$^{-1}$, while the SO emission
is detected in a narrower velocity range between $\sim$5.9~km~s$^{-1}$ to 
$\sim$7.5~km~s$^{-1}$, i.e., SO is mainly
coming from the 6.5~km~s$^{-1}$ component. The intensity peak at 6.5~km~s$^{-1}$ 
is shifted $\sim$5$''$ west from the peak at 7.0~km~s$^{-1}$. It is interesting to note that DCO$^+$   
presents an emission peak toward B1b-N at a velocity of $\sim$7.5~km~s$^{-1}$ 
in agreement with the N$_2$H$^+$ and NH$_2$D observations by \cite{Daniel13}.
Recent interferometric observations revealed the kinematical structure of this 
region at scales of $\sim$500-1000~AU \citep{Huang13,Gerin15}. \citet{Huang13} 
detected intense emission of the  N$_2$D$^+$ 3$\rightarrow$2 lines toward the 
two protostellar cores B1b-N and B1b-S using the SMA. 
The N$_2$D$^+$ emission is centered at $\sim$7.0$\pm$0.2~km~s$^{-1}$ toward 
B1b-N and at $\sim$6.5$\pm$0.2~km~s$^{-1}$  toward B1b-S, confirming the existence 
of the two velocity components. 

In the 
cases of H$_2$$^{13}$CO, H$_2$S, HOCO$^+$, and in one transition of 
CCS, OCS, and SO$_2$ (see Table B.1), we detect wings at redshifted velocities. 
Compact outflows are detected toward each protostellar core with 
wings in the 0 to 15~km~s$^{-1}$ velocity range \citep{Gerin15}. 
In Fig.~\ref{fig6} we overlay the 30 m beam at 3 mm and 1 mm on
the interferometric H$_2$CO map. The B1b-N outflow and the red wing of the B1b-S outflow 
lie within our beam at 3 mm, but the whole B1b-N outflow and a fraction of the red
lobe of the B1b-S outflow are missed in 
the 2 mm and 1 mm observations. The emission at the outflow velocities
is not taken into account in our calculations because we are only interested in 
the two velocity components associated with the core emission (see Sect.~5 for details).

\begin{table*}
\caption{Column density calculations}

\label{tab_column}
\begin{tabular}{llllll|lll} \hline \hline
Molecule     &  \multicolumn{5}{c|}{LTE}     & \multicolumn{3}{c}{LVG$^1$}  \\
             &  \multicolumn{2}{c}{6.5~km~s$^{-1}$}                             &      \multicolumn{2}{c}{7.0~km~s$^{-1}$}  
 &      \multicolumn{1}{c|}{Total}  & \multicolumn{1}{c}{6.5~km~s$^{-1}$}                             &      \multicolumn{1}{c}{7.0~km~s$^{-1}$}  
 &      \multicolumn{1}{c}{Total} \\
                  &   T$_{rot}$(K)          &  N$_{X}$(cm$^{-2}$)                     &     T$_{rot}$(K)   & N$_X$(cm$^{-2}$)  &  N$_X$(cm$^{-2}$)  &
 N$_{X}$(cm$^{-2}$)          &   N$_{X}$(cm$^{-2}$)          &  N$_{X}$(cm$^{-2}$)          \\ \hline \hline
%
%
\multicolumn{6}{c|}{\it Ions} & \multicolumn{3}{c}{\it Ions}\\
HC$^{18}$O$^+$    &  4.4$_{-0.2}^{+0.2}$   &  1.7$_{-0.1}^{+0.1}$$\times$10$^{11}$    &   4.2$_{-0.7}^{+0.4}$   &  1.5$_{-0.3}^{+0.2}$$\times$10$^{11}$ &
3.2$\times$10$^{11}$  &   2.0$_{-0.1}^{+0.1}$$\times$10$^{11}$  &  1.4$_{-0.3}^{+0.2}$$\times$10$^{11}$  &  3.4$\times$10$^{11}$  \\
D$^{13}$CO$^+$    &  7.8$_{-0.5}^{+0.5}$   &  1.2$_{-0.2}^{+0.2}$$\times$10$^{11}$    &   4.5$_{-0.2}^{+0.2}$   &  3.9$_{-0.5}^{+0.5}$$\times$10$^{11}$ &
5.1$\times$10$^{11}$  &  2.0$_{-0.1}^{+0.1}$$\times$10$^{11}$ &   4.5$_{-0.1}^{+0.1}$$\times$10$^{11}$ & 6.5$\times$10$^{11}$  \\  
HOCO$^+$          &  11.9$_{-0.3}^{+0.3}$  &  5.2$_{-0.3}^{+0.3}$$\times$10$^{11}$    &   10.7$_{-1.0}^{+1.3}$  &  2.4$_{-0.7}^{+1.0}$$\times$10$^{11}$  &
7.6$\times$10$^{11}$ &  5.2$_{-0.2}^{+0.2}$$\times$10$^{11}$  &  3.0$_{-0.2}^{+0.2}$$\times$10$^{11}$ & 8.2$\times$10$^{11}$ \\
DOCO$^+$          &  11$^a$                &  7.0$_{-0.9}^{+0.9}$$\times$10$^{10}$    &   11$^a$                &  4.0$_{-2.0}^{+2.0}$$\times$10$^{10}$  &
1.1$\times$10$^{11}$ &   9.0$_{-1.1}^{+1.1}$$\times$10$^{10}$  &  5.0$_{-2.0}^{+2.0}$$\times$10$^{10}$ &  1.1$\times$10$^{11}$ \\
SO$^+$            &  9.0$_{-0.7}^{+0.8}$   &  7.6$_{-1.6}^{+2.0}$$\times$10$^{11}$    &   8.3$_{-1.4}^{+2.2}$   &  4.8$_{-2.2}^{+4.1}$$\times$10$^{11}$  & 
1.2$\times$10$^{12}$ &       &         &         \\
HCS$^+$           &  10.6$_{-3.0}^{+6.5}$  &  2.1$_{-1.3}^{+3.6}$$\times$10$^{11}$    &   7.5$_{-0.6}^{+0.5}$     &  6.2$_{-0.2}^{+0.2}$$\times$10$^{11}$ & 
8.3$\times$10$^{11}$  &   5.5$_{-0.1}^{+0.1}$$\times$10$^{11}$  &   5.5$_{-0.1}^{+0.1}$$\times$10$^{11}$  &  1.1$\times$10$^{12}$    \\
DCS$^+$           &  10$^b$                &  5.0$_{-1.0}^{+1.0}$$\times$10$^{10}$    &                           &   & 
5.0$\times$10$^{10}$  &  4.5$_{-1.1}^{+1.1}$$\times$10$^{10}$ &           &  4.5$\times$10$^{10}$   \\
HCNH$^+$           &  29.0$_{-12}^{+50}$  &  5.1$_{-0.7}^{+4.1}$$\times$10$^{12}$    &   8.9$_{-2.9}^{+5.3}$     &  3.8$_{-0.5}^{+1.5}$$\times$10$^{12}$ & 
8.9$\times$10$^{12}$  &   3.7$_{-0.4}^{+0.4}$$\times$10$^{12}$  &  4.0$_{-0.4}^{+0.4}$$\times$10$^{12}$  & 7.7$\times$10$^{12}$ \\
H$_2$COH$^+$       &                     &                                           &   10$^{a}$     &  2.0$_{-1.0}^{+1.0}$$\times$10$^{11}$ & 
 2.0$_{-1.0}^{+1.0}$$\times$10$^{11}$  &       &          &  \\
CO$^+$       &                     &                                           &   10$^{a}$     &  $<$6.0$\times$10$^{10}$ &      
$<$6.0$\times$10$^{10}$  &        &        &    \\
HOC$^+$             &                     &                                           &   10$^{a}$     &  $<$2.0$\times$10$^{10}$ & 
$<$2.0$\times$10$^{10}$ &     &       &   \\ \hline
%
%
\multicolumn{6}{c|}{\it Neutrals} & \multicolumn{3}{c}{\it Neutrals}\\
$^{13}$C$^{18}$O  &  5.5$_{-0.9}^{+1.3}$   &  2.4$_{-0.2}^{+0.2}$$\times$10$^{13}$    &   10.9$_{-1.4}^{+0.9}$  &  5.2$_{-0.1}^{+0.1}$$\times$10$^{13}$ &
7.6$\times$10$^{13}$ & 3.0$_{-0.2}^{+0.2}$$\times$10$^{13}$  &  6.0$_{-0.3}^{+0.3}$$\times$10$^{13}$ &   9.0$\times$10$^{13}$  \\   
o-H$_2$$^{13}$CO  &  5.5$_{-0.6}^{+0.8}$   &  6.1$_{-1.5}^{+2.1}$$\times$10$^{11}$    &   5.2$_{-0.3}^{+0.9}$   &  4.9$_{-1.9}^{+3.2}$$\times$10$^{11}$  &
1.1$\times$10$^{12}$ &  1.1$_{-0.6}^{+0.6}$$\times$10$^{12}$ &   1.3$_{-0.7}^{+0.7}$$\times$10$^{12}$  & 2.4$\times$10$^{12}$ \\
$^{13}$CS         &  7.1$_{-1.1}^{+1.6}$   &  7.8$_{-5.7}^{+2.0}$$\times$10$^{11}$    &   3.2$_{-0.7}^{+0.4}$   &  3.1$_{-1.6}^{+3.2}$$\times$10$^{12}$  &
3.9$\times$10$^{12}$  &  1.3$_{-0.1}^{+0.1}$$\times$10$^{12}$  &  9.0$_{-0.3}^{+0.3}$$\times$10$^{11}$  &   2.2$_{-0.1}^{+0.1}$$\times$10$^{12}$ \\   
S$^{18}$O         &  6.8$_{-0.6}^{+0.6}$   &  1.6$_{-0.5}^{+0.6}$$\times$10$^{12}$    &   8.2$_{-1.4}^{+2.0}$   &  0.6$_{-0.3}^{+0.4}$$\times$10$^{12}$  &
2.3$\times$10$^{12}$ &  2.2$_{-0.1}^{+0.1}$$\times$10$^{12}$  & 8.0$_{-0.3}^{+0.3}$$\times$10$^{11}$ &  3.0$\times$10$^{12}$  \\         
$^{33}$SO         &  12.3$_{-4.0}^{+7.3}$  &  1.6$_{-0.6}^{+1.0}$$\times$10$^{12}$    &   18.9$_{-8.3}^{+71}$   &  0.7$_{-0.4}^{+0.7}$$\times$10$^{12}$ &
2.3$\times$10$^{12}$  &  3.8$_{-0.1}^{+0.1}$$\times$10$^{12}$  &  6.0$_{-2.0}^{+2.0}$$\times$10$^{11}$  &   4.4$\times$10$^{12}$  \\  
$^{34}$SO$_2$     &  5.8$_{-1.2}^{+2.0}$   &  1.1$_{-0.5}^{+1.0}$$\times$10$^{12}$     &           &           &   
& 1.1$_{-0.1}^{+0.1}$$\times$10$^{12}$   &                &    \\
SO$_2$            &  10.1$_{-1.0}^{+1.0}$  &  7.9$_{-2.0}^{+3.0}$$\times$10$^{12}$     &  10.2$_{-1.2}^{+2.0}$  &   2.9$_{-0.9}^{+1.3}$$\times$10$^{12}$  &
1.1$\times$10$^{13}$  &  1.6$_{-0.1}^{+0.1}$$\times$10$^{13}$   &   4.5$_{-0.1}^{+0.1}$$\times$10$^{12}$  & 2.0$\times$10$^{13}$  \\
OC$^{34}$S        &  8.7$_{-0.5}^{+0.6}$   &  1.2$_{-0.2}^{+0.2}$$\times$10$^{12}$    &                         &                                        &
  &  1.0$_{-0.3}^{+0.3}$$\times$10$^{12}$ &            &        \\ 
OCS               & 11.7$_{-4.0}^{+13.0}$    &  1.5$_{-0.6}^{+0.9}$$\times$10$^{13}$    &  20.4$_{-1.3}^{+1.4}$  & 4.9$_{-0.6}^{+0.6}$$\times$10$^{12}$   &
1.9$\times$10$^{13}$  & 1.7$_{-0.1}^{+0.1}$$\times$10$^{13}$       &  8.0$_{-4.0}^{+4.0}$$\times$10$^{11}$    &  1.7$\times$10$^{13}$  \\
H$_2$$^{13}$CS$^c$     &   7.4                  &  1.6$_{-0.3}^{+0.3}$$\times$10$^{11}$     &            &          &   
&                 &                        &          \\
CCS               &  8.3$_{-0.6}^{+0.7}$   &  5.9$_{-1.7}^{+2.4}$$\times$10$^{12}$   &    8.2$_{-0.6}^{+0.7}$  &  3.6$_{-1.0}^{+1.3}$$\times$10$^{12}$  &
9.5$\times$10$^{12}$  &   9.3$_{-0.1}^{+0.1}$$\times$10$^{12}$   & 5.2$_{-0.1}^{+0.1}$$\times$10$^{12}$ & 1.4$\times$10$^{13}$  \\
CC$^{34}$S        &    &   &   &   &  & 3.5$_{-0.7}^{+0.7}$$\times$10$^{11}$   &   2.5$_{-0.7}^{+0.7}$$\times$10$^{11}$ &   6.0$\times$10$^{11}$  \\
CCCS              &   &   &  &  &    &  &   &  \\ 
C$^{13}$CCS       &   &   &  &  &    &  &  &  \\   
o-H$_2$S    &  10$^d$  &  7.0$_{-0.2}^{+0.2}$$\times$10$^{12}$  &  10 &  3.0$_{-0.2}^{+0.2}$$\times$10$^{12}$  &    1.0$\times$10$^{13}$   &
 &  &  \\
NS          &  10$^d$   &  5.4$_{-0.2}^{+0.2}$$\times$10$^{12}$ &  10 &  9.0$_{-0.2}^{+0.2}$$\times$10$^{12}$  &    1.4$\times$10$^{13}$   &
 &  &  \\
%
%
%
%
\hline \hline
\end{tabular}

\noindent
$^1$ LVG calculations assuming T$_k$=12~K and n(H$_2$)=10$^5$~cm$^{-3}$; 
$^a$ Assuming the rotation temperature of HOCO$^+$;
$^b$ Assuming the rotation temperature of HCS$^+$;
$^c$ In this case the ortho and para species are treated together (ortho-to-para(OTP)=3) and we assumed the rotation temperature calculated for H$_2$CS  by \cite{Marcelino05}; 
$^d$ Reasonable guess for the rotation temperature.
\end{table*}

\begin{table*}
\caption{Fractional abundances and abundance ratios}

\label{tab_abundance}
\begin{tabular}{llll|lll} \hline \hline
Molecule     &  \multicolumn{3}{c|}{LTE}     & \multicolumn{3}{c}{LVG$^1$}  \\
             &  \multicolumn{1}{c}{6.5~km~s$^{-1}$}     &      \multicolumn{1}{c}{7.0~km~s$^{-1}$} &    \multicolumn{1}{c|}{Total} &
 \multicolumn{1}{c}{6.5~km~s$^{-1}$}     &      \multicolumn{1}{c}{7.0~km~s$^{-1}$} &    \multicolumn{1}{c}{Total}\\ \hline 
\hline
%
\multicolumn{4}{c|}{\it Ions} & \multicolumn{3}{c}{\it Ions} \\
X(HC$^{18}$O$^+$)$\times$550    &  3.9$\times$10$^{-9}$    &   1.8$\times$10$^{-9}$  &   2.5$\times$10$^{-9}$ &  4.6$\times$10$^{-9}$ &  1.7$\times$10$^{-9}$ &  
{\bf 2.7$\times$10$^{-9}$} \\
X(D$^{13}$CO$^+$)$\times$60     &  2.7$\times$10$^{-10}$    &   4.7$\times$10$^{-10}$  &   4.0$\times$10$^{-10}$ &  4.6$\times$10$^{-10}$ & 5.4$\times$10$^{-10}$ &  
{\bf 5.1$\times$10$^{-10}$}\\  
X(HOCO$^+$)                     &  2.0$\times$10$^{-11}$   &   4.8$\times$10$^{-12}$  &  1.0$\times$10$^{-11}$ & 2.0$\times$10$^{-11}$ &  6.0$\times$10$^{-12}$ 
& {\bf 1.1$\times$10$^{-11}$} \\
X(DOCO$^+$)                     &  2.9$\times$10$^{-12}$   &   8.0$\times$10$^{-13}$  &  1.4$\times$10$^{-12}$ & 3.6$\times$10$^{-12}$  & 1.0$\times$10$^{-12}$ &
 {\bf 1.4$\times$10$^{-12}$} \\  
X(SO$^+$)                       &  3.1$\times$10$^{-11}$   &   9.6$\times$10$^{-12}$  &  {\bf 1.6$\times$10$^{-11}$} &    &    &  \\
X(HCS$^+$)                      &  8.1$\times$10$^{-12}$   &   1.2$\times$10$^{-11}$  &  1.1$\times$10$^{-11}$ &  2.1$\times$10$^{-11}$ &  1.1$\times$10$^{-11}$ &
{\bf 1.4$\times$10$^{-11}$} \\
X(DCS$^+$)                      &  1.9$\times$10$^{-12}$   &                          &  6.6$\times$10$^{-13}$ &  1.7$\times$10$^{-12}$ & & {\bf 5.9$\times$10$^{-13}$}  \\
X(HCNH$^+$)                     &  2.0$\times$10$^{-10}$   &   7.6$\times$10$^{-11}$  &  1.2$\times$10$^{-10}$ &  1.4$\times$10$^{-10}$ & 8.0$\times$10$^{-11}$ &
{\bf 1.0$\times$10$^{-10}$} \\ 
X(H$_2$COH$^+$)                 &                          &   4.0$\times$10$^{-12}$  &  {\bf 2.6$\times$10$^{-12}$} & & &\\ 
X(CO$^+$)                       &                          &                          &  {\bf $<$7.9$\times$10$^{-13}$}  & & &\\     
X(HOC$^+$)                      &                          &                          &  {\bf $<$2.6$\times$10$^{-13}$}  & & &\\
X(NNH$^+$)                      &                          &                          &  {\bf 1.4$\times$10$^{-9}$$^*$}  & & & \\ 
X(NND$^+$)                      &                          &                          &  {\bf 5.0$\times$10$^{-10}$$^*$} & & &\\ \hline 
\multicolumn{4}{c|}{\it Neutrals} & \multicolumn{3}{c}{\it Neutrals} \\
%
%
X($^{13}$C$^{18}$O)          &  1.0$\times$10$^{-9}$    &   1.0$\times$10$^{-9}$  &   1.0$\times$10$^{-9}$ &  1.2$\times$10$^{-9}$ &  1.2$\times$10$^{-9}$ &  {\bf 1.2$\times$10$^{-9}$}  \\   
X(H$_2$$^{13}$CO)$\times$60$^a$     &  1.9$\times$10$^{-9}$    &   7.8$\times$10$^{-10}$   &  1.1$\times$10$^{-9}$ &  3.4$\times$10$^{-9}$ &  2.1$\times$10$^{-9}$ &
 {\bf 2.5$\times$10$^{-9}$} \\
X($^{13}$CS)$\times$60          &  1.8$\times$10$^{-9}$    &   3.7$\times$10$^{-9}$   &  3.1$\times$10$^{-9}$  &  3.0$\times$10$^{-9}$  &  1.1$\times$10$^{-9}$ &
 {\bf 1.7$\times$10$^{-9}$} \\
X(S$^{18}$O)$\times$550         &  3.7$\times$10$^{-8}$    &   6.3$\times$10$^{-9}$   &  1.8$\times$10$^{-8}$ &  {\bf 5.0$\times$10$^{-8}$}  &  {\bf 9.0$\times$10$^{-9}$}  &
{\bf 2.2$\times$10$^{-8}$}  \\  
X($^{33}$SO)$\times$135         &  9.0$\times$10$^{-9}$    &   1.8$\times$10$^{-9}$   &  4.0$\times$10$^{-9}$ &  1.9$\times$10$^{-8}$  & 1.6$\times$10$^{-9}$  &
 7.9$\times$10$^{-9}$  \\
X($^{34}$SO$_2$)$\times$22.5  &   1.0$\times$10$^{-9}$    &      &   &  1.0$\times$10$^{-9}$ &  &  \\
X(SO$_2$)                     &   3.3$\times$10$^{-10}$   &   5.6$\times$10$^{-11}$  &   1.4$\times$10$^{-10}$  & 
{\bf 6.7$\times$10$^{-10}$}  &   {\bf 9.0$\times$10$^{-11}$}  &   {\bf 2.6$\times$10$^{-10}$}   \\        
X(OC$^{34}$S)$\times$22.5    &  1.1$\times$10$^{-9}$  &         &          &
9.4$\times$10$^{-10}$        &                        &       \\
X(OCS)                       &  6.2$\times$10$^{-10}$  &    9.4$\times$10$^{-11}$  &    2.4$\times$10$^{-10}$  &
  {\bf 7.0$\times$10$^{-10}$}      &   {\bf 1.5$\times$10$^{-11}$}  &   {\bf 2.2$\times$10$^{-10}$} \\     
X(H$_2$$^{13}$CS)$\times$60   &  4.0$\times$10$^{-10}$  &         &              &
 &   &  \\
X(CCS)    &  2.4$\times$10$^{-10}$  &  6.9$\times$10$^{-11}$  &  1.2$\times$10$^{-10}$  &
 3.9$\times$10$^{-10}$  &  1.0$\times$10$^{-10}$ &  {\bf 1.8$\times$10$^{-10}$} \\  
X(H$_2$S)$^a$  &  3.9$\times$10$^{-10}$  &  7.7$\times$10$^{-11}$  &  {\bf 1.7$\times$10$^{-10}$}  & & & \\ 
X(NS)  &  2.2$\times$10$^{-10}$ &   1.7$\times$10$^{-10}$  &   {\bf 1.8$\times$10$^{-10}$}  &  &  &  \\
X(HCN)                          &                          &                          &  {\bf 1.2$\times$10$^{-8}$$^*$}  & & &\\ \hline 
%
%
\multicolumn{4}{c|}{\it Ion/Neutral and Ion/Ion} & \multicolumn{3}{c}{\it Ion/Neutral and Ion/Ion} \\
$\frac{N(HCO^+)}{N(CO)}$           &  1.1$\times$10$^{-4}$         &   4.8$\times$10$^{-5}$  &   6.9$\times$10$^{-5}$ &  1.1$\times$10$^{-4}$ &   4.3$\times$10$^{-5}$ &   
{\bf 6.8$\times$10$^{-5}$}   \\
$\frac{N(DCO^+)}{N(HCO^+)}$        &  0.07                         &   0.26                  &   0.15          &  0.1 &   0.32 &  {\bf 0.19}     \\
$\frac{N(DCO^+)}{N(CO)}$           &  9.0$\times$10$^{-6}$         &   1.2$\times$10$^{-5}$  &   1.1$\times$10$^{-5}$ &  1.1$\times$10$^{-5}$ &  1.4$\times$10$^{-5}$ & 
{\bf 1.2$\times$10$^{-5}$} \\
$\frac{N(HOCO^+)}{N(HCO^+)}$       &  0.005                        &   0.003                 &   0.004               & 0.004 & 0.003 & {\bf 0.004} \\
$\frac{N(SO^+)}{N(SO)}$            &  8.0$\times$10$^{-4}$         &   0.0014                &   {\bf 9.5$\times$10$^{-4}$} &    &  &   \\
$\frac{N(HCS^+)}{N(CS)}$           &  0.004                        &   0.003                 &   0.0035  & 0.007 & 0.01 & {\bf 0.008} \\
$\frac{N(DCS^+)}{N(HCS^+)}$        &  0.24                         &                         &   0.06    & 0.08 & & {\bf 0.04} \\
$\frac{N(HCS^+)}{N(HCO^+)}$        &  0.003                        &   0.007                 &   0.004   & 0.004 & 0.006 &  {\bf 0.005}   \\
$\frac{N(HCNH^+)}{N(HCN)}$         &                               &                         &   0.01   & & & {\bf 0.008} \\
$\frac{N(H_2COH^+)}{N(H_2CO)}$     &                               &   0.005                 &   {\bf 0.002}  & &  &  \\
$\frac{N(CO^+)}{N(HCO^+)}$         &        &    &   {\bf $<$3$\times$10$^{-4}$} &   &   &  \\   
$\frac{N(HOC^+)}{N(HCO^+)}$        &        &    &   {\bf $<$1$\times$10$^{-4}$} &   &   &   \\ \hline 
%
%
\multicolumn{4}{c|}{\it Neutral/Neutral}  & \multicolumn{3}{c}{\it Neutral/Neutral} \\
$\frac{N(SO)}{N(CS)}$       &  19      &   1.8    &   5   & 17 & 8 & {\bf 13} \\
$\frac{N(SO_2)}{N(SO)}$     &  0.03   &   0.008  &  0.008 & 0.01 & 0.009 & {\bf 0.01} \\   
$\frac{N(CCS)}{N(CS)}$      &  0.13    &   0.02   &  0.04  & 0.12  & 0.09  &  {\bf 0.11} \\ 
$\frac{N(H_2CS)}{N(CS)}$    &  {\bf0.20}    &          &        &     &   &   \\
$\frac{N(OCS)}{N(CS)}$      &  0.57    &   0.03  &  0.08  & 0.22 & 0.01 & {\bf 0.13} \\ 
\hline \hline 
\end{tabular}

\noindent
$^1$ LVG calculations assuming T$_k$=12~K and n(H$_2$)=10$^5$~cm$^{-3}$; 
numbers in boldface are used in Fig.~\ref{fig7} to Fig.~\ref{fig11} to compare with observations.
$^*$ From Daniel et al. (2013); 
$^a$ From Table~2 assuming OTP=3.
\end{table*}

\begin{figure*}
\hspace*{-0.0cm}
\includegraphics[scale=0.80,angle=0]{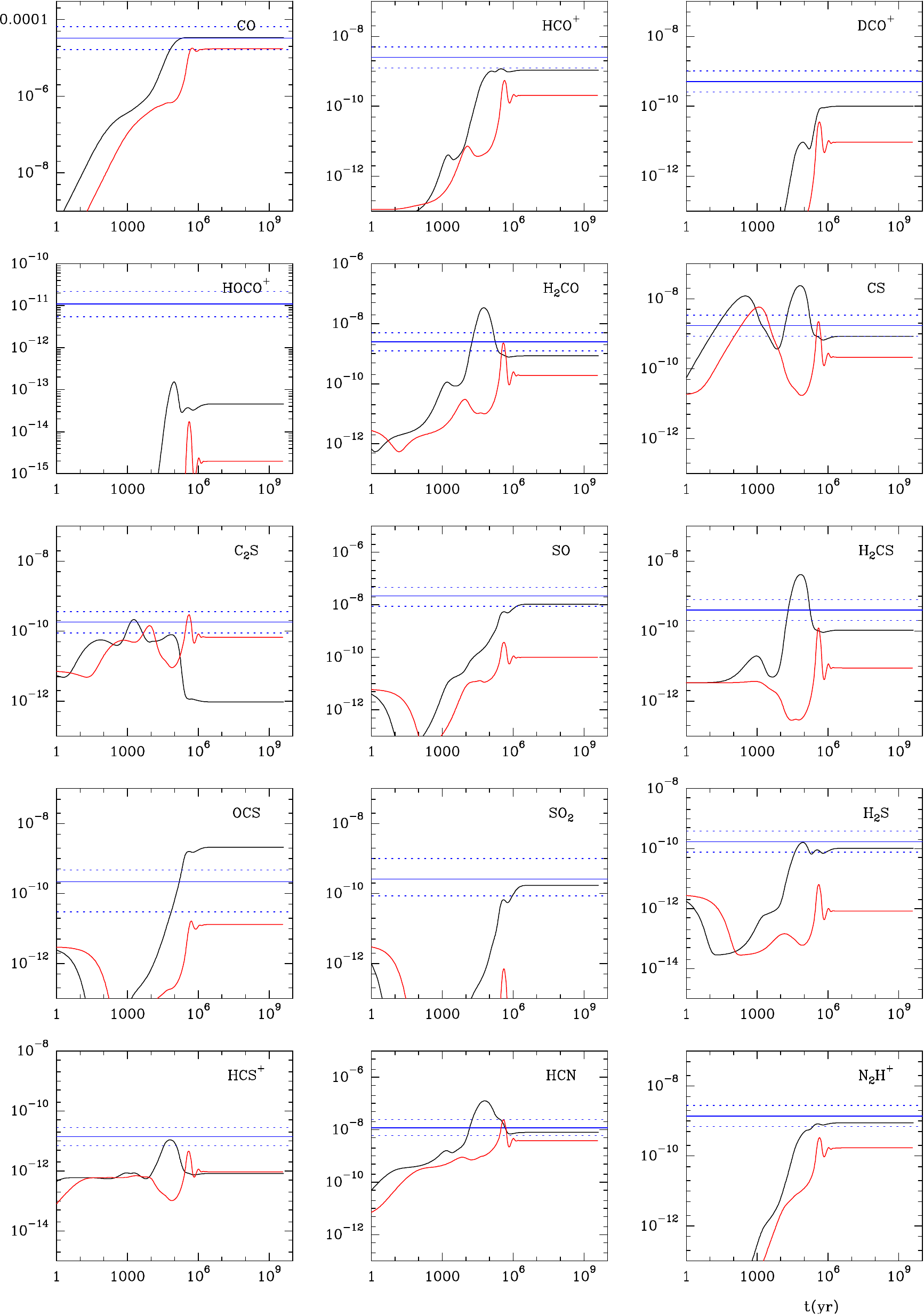}
   \caption{\label{fig7}
Results of our best-fit model ($\zeta_{H_2}$=4.0$\times$10$^{-17}$~s$^{-1}$, O/H=3.0~10$^{-5}$, 
N/H=6.4~10$^{-5}$, C/H=1.7~10$^{-5}$, and S/H=6.0~10$^{-7}$).
We assume a density of n(H$_2$)=10$^5$~cm$^{-3}$ and T$_k$=12~K. The thick and dashed blue lines indicate the observed 
values and their uncertainty, which are considered to be a factor of 2 for most molecules (see text), respectively. 
In the case of SO, SO$_2$, and OCS, their abundances change by more than
a factor of 2 between the 6.5~km~s$^{-1}$ and 7.0~km~s$^{-1}$ velocity components. 
In the panels corresponding to these species, the dashed lines indicate the abundance for 
each component separately. In red we show model calculations for the same conditions and  
n(H$_2$)=10$^4$~cm$^{-3}$. 
}
 \end{figure*}

\begin{figure*}
\hspace*{-0.0cm}
\includegraphics[scale=0.80,angle=0]{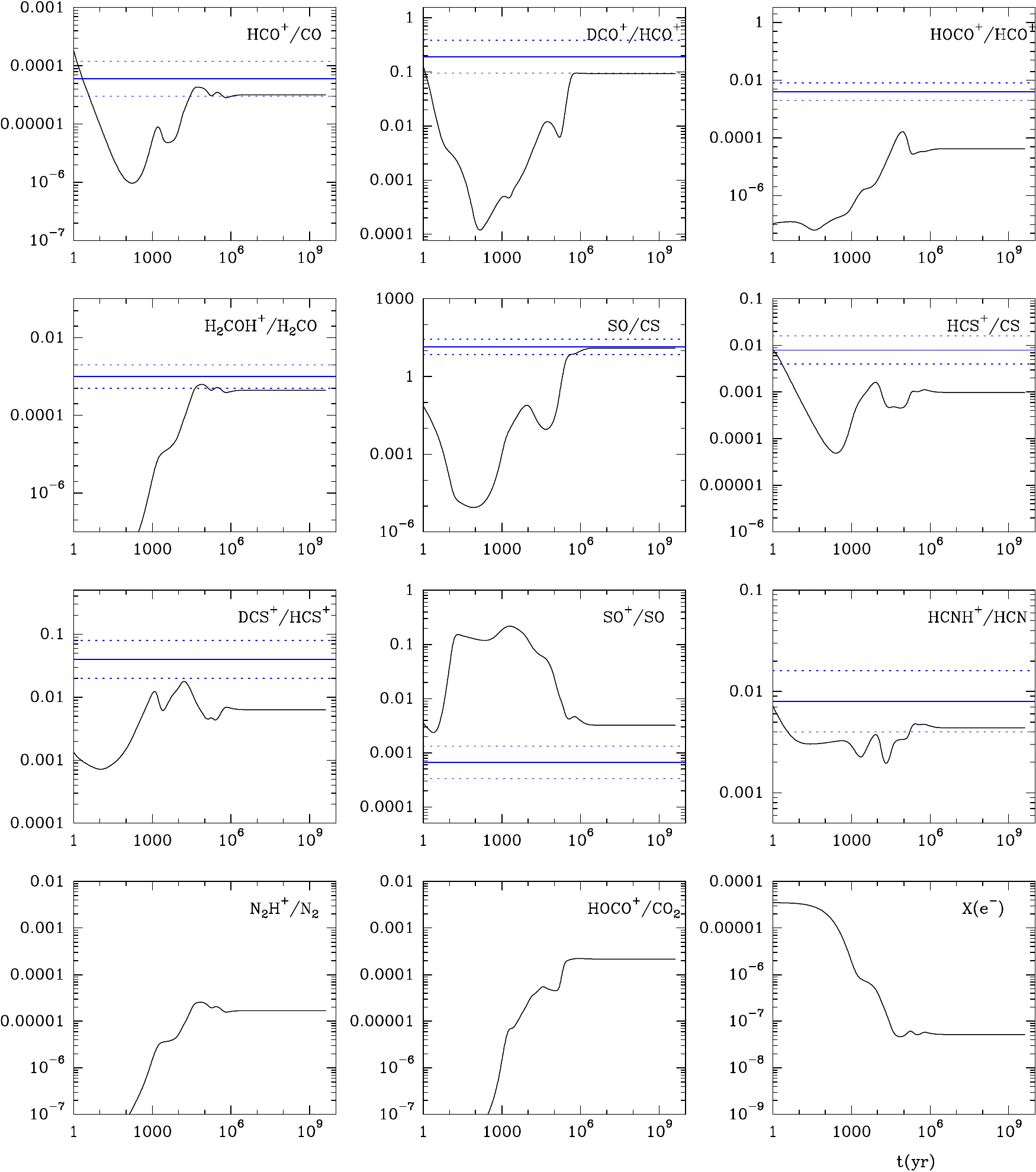}
   \caption{\label{fig8}
Results of our best-fit model ($\zeta_{H_2}$=4.0$\times$10$^{-17}$~s$^{-1}$, O/H=3.0~10$^{-5}$, 
N/H=6.4~10$^{-5}$, C/H=1.7~10$^{-5}$, and S/H=6.0~10$^{-7}$).
We assume a density of n(H$_2$)=10$^5$~cm$^{-3}$ and T$_k$=12~K. The thick and dashed blue lines indicate the observed 
values and their uncertainties. 
}
 \end{figure*}

\section{Column density calculations}

In order to investigate abundance variations between the two velocity components, 
we carried out a separate analysis for each. The spectral resolution of our observations is different for the 3 mm, 2 mm and 1 mm bands. 
To carry out a uniform analysis, we degraded the velocity resolution to a common value of 0.25~km~s$^{-1}$ and then 
fitted Gaussians to each one. In the fitting, we fixed the central velocities to 6.5~km~s$^{-1}$ and $\sim$7.0~km~s$^{-1}$, 
and the linewidths to $\Delta$v=0.7~km~s$^{-1}$ and 1.0~km~s$^{-1}$, respectively. The fitting returns the integrated intensity for
each component. This procedure gives an excellent fit for most lines and 
neglects possible line wings with velocities $>$7.5~km~s$^{-1}$. The 
final integrated intensities are shown in Table B.1.
We are aware that this simple method is not able to cleanly separate the contribution from each component. 
There is some unavoidable velocity overlap between the two components along the line of sight and  
our spatial resolution is limited (see Fig~\ref{fig4} and Fig~\ref{fig5}). 
In spite of that, our simple method help to detect possible chemical gradients.

Several isotopologues of each molecule were detected as expected taking the large extinction in this core into
account  (A$_v$$\sim$76~mag).
As zero-order approximation, we derived the column densities and fractional abundances of the different species using the rotation
diagram technique. This technique gives a good estimate of the total number of molecules within our beam as far as the
observed lines are optically thin. To be sure of this requirement, we selected the rarest isotopologue of each species. Furthermore,
we assumed that the emission is filling the beam. To calculate the column densities and abundances of the main isotopologues, we 
adopted $^{12}$C/$^{13}$C=60, $^{16}$O/$^{18}$O=550, $^{32}$S/$^{34}$S=22.5, and $^{34}$S/$^{33}$S=6 \citep{Chin96,Wilson99}.

In Table~\ref{tab_column} we show the results of our rotational diagram calculations. 
The errors in the rotation temperatures and column densities are those
derived from the least squares fitting. There are, however, other kind
of uncertainties that are not considered in these numbers. One source of uncertainty
is the unknown beam filling factor. Since the beam is dependent on the
frequency, this could bias the derived rotation temperature. 
To estimate the error from this uncertainty, we calculated
the column densities with an alternative method. \citet{Daniel13} modeled the density
structure of the Barbard~1b core on the basis of dust continuum at 350~$\mu$m and 1.2~mm.
These authors obtained that the mean density in the $\sim$29$\arcsec$ beam is 
n(H$_2$)$\approx$10$^5$~cm$^{-3}$. \cite{Lis10} derived a gas kinetic temperature of T$_K$ = 12~K 
from NH$_3$ observations. We adopted these physical conditions to derive the molecular
column densities by fitting the integrated intensity of the lowest energy transition 
using the LVG code MADEX \citep{Cernicharo12a}.
The results are shown in Table~\ref{tab_column}. Column densities derived using the rotation diagram technique and
LVG calculations agree within a factor of 2 for all the compounds. We consider that our estimates
are accurate within this factor.

\begin{figure*}
\hspace*{-0.0cm}
\includegraphics[scale=0.80,angle=0]{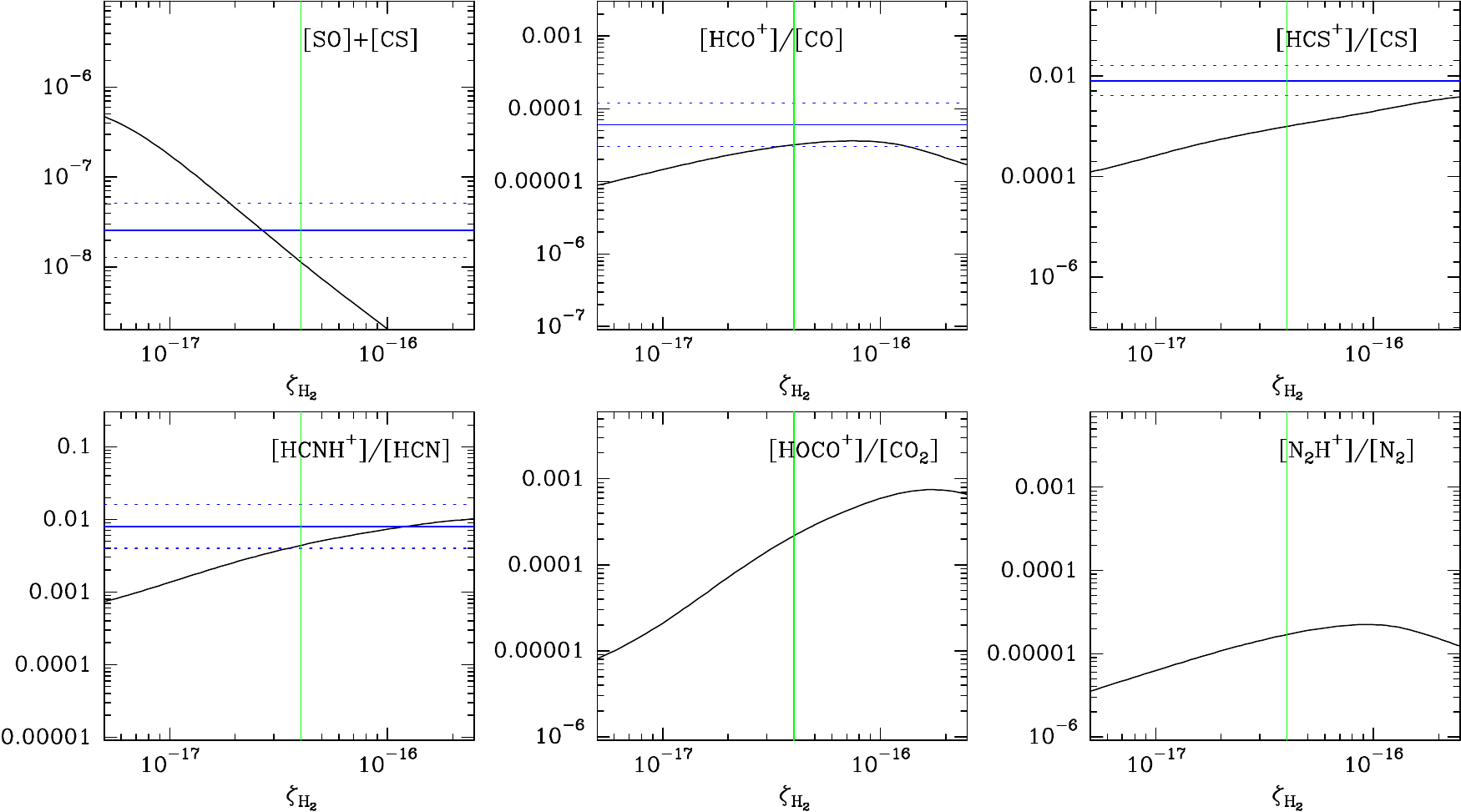}
   \caption{\label{fig9}
Steady-state calculations of fractional abundances and abundance ratios. All the input parameters
except $\zeta_{H_2}$ are the same as in Fig.~\ref{fig7}. Horizontal blue lines correspond to the 
observed values and their uncertainty that is considered a factor of 2. 
Vertical green lines indicate the estimated value of the cosmic ray ionization rate,
$\zeta_{H_2}$=4$\times$10$^{-17}$~s$^{-1}$.
}
 \end{figure*}

\begin{figure*}
\hspace*{-0.0cm}
\includegraphics[scale=0.80,angle=0]{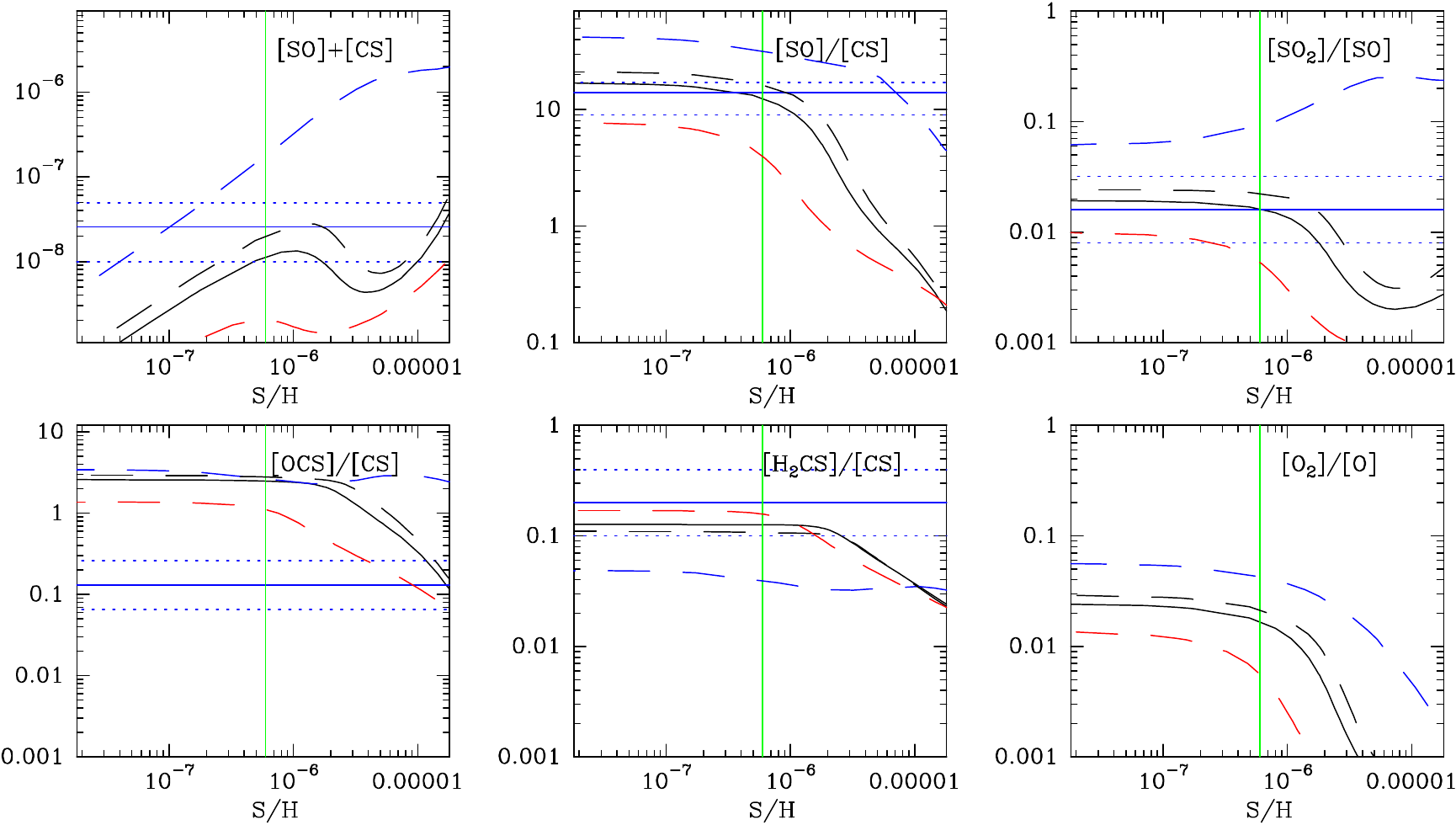}
   \caption{\label{fig10}
   Steady-state calculations of fractional abundances and abundance ratios. All the input parameters
except S/H are the same as in Fig.~\ref{fig7}. Horizontal blue lines correspond to the 
observed values and their uncertainty. Vertical green lines indicate the estimated value of the sulfur elemental abundance,
S/H=6$\times$10$^{-7}$. The solid black line represents our best model fit, $\zeta_{H_2}$=4$\times$10$^{-17}$~s$^{-1}$. The dashed blue, black, and red lines correspond to $\zeta_{H_2}$=1, 4 and 10$\times$10$^{-17}$~s$^{-1}$, 
respectively.
}
 \end{figure*}

\begin{figure}
\hspace*{-0.0cm}
\includegraphics[scale=0.65,angle=0]{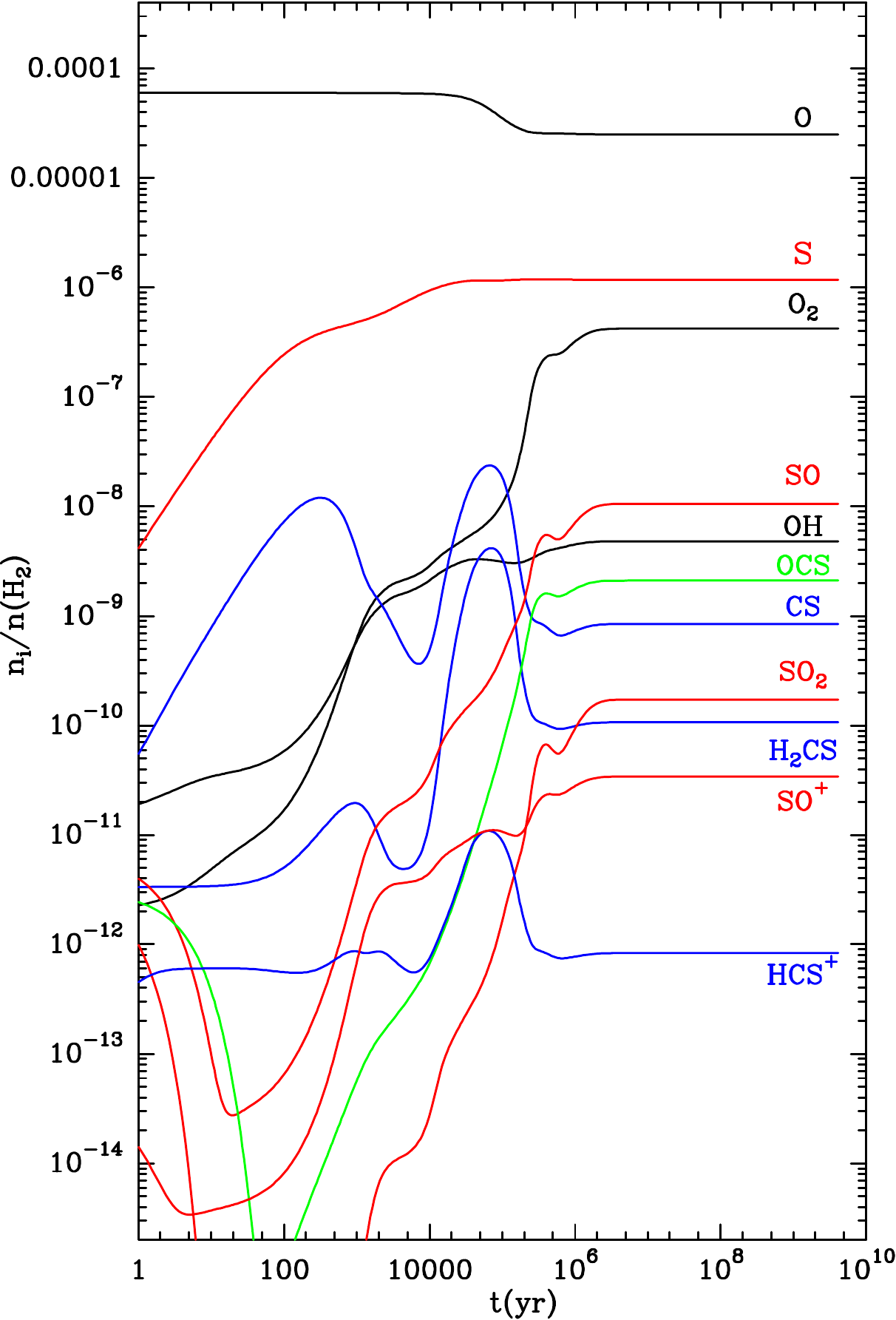}
   \caption{\label{fig11}
Predicted fractional abundances using the same parameters as in Fig~\ref{fig7}. 
The abundances of SO, SO$_2$ and SO$^+$, and O$_2$ dramatically increase after 1~Myr.
}
 \end{figure}

\begin{figure*}
\hspace*{-0.0cm}
\includegraphics[scale=0.80,angle=0]{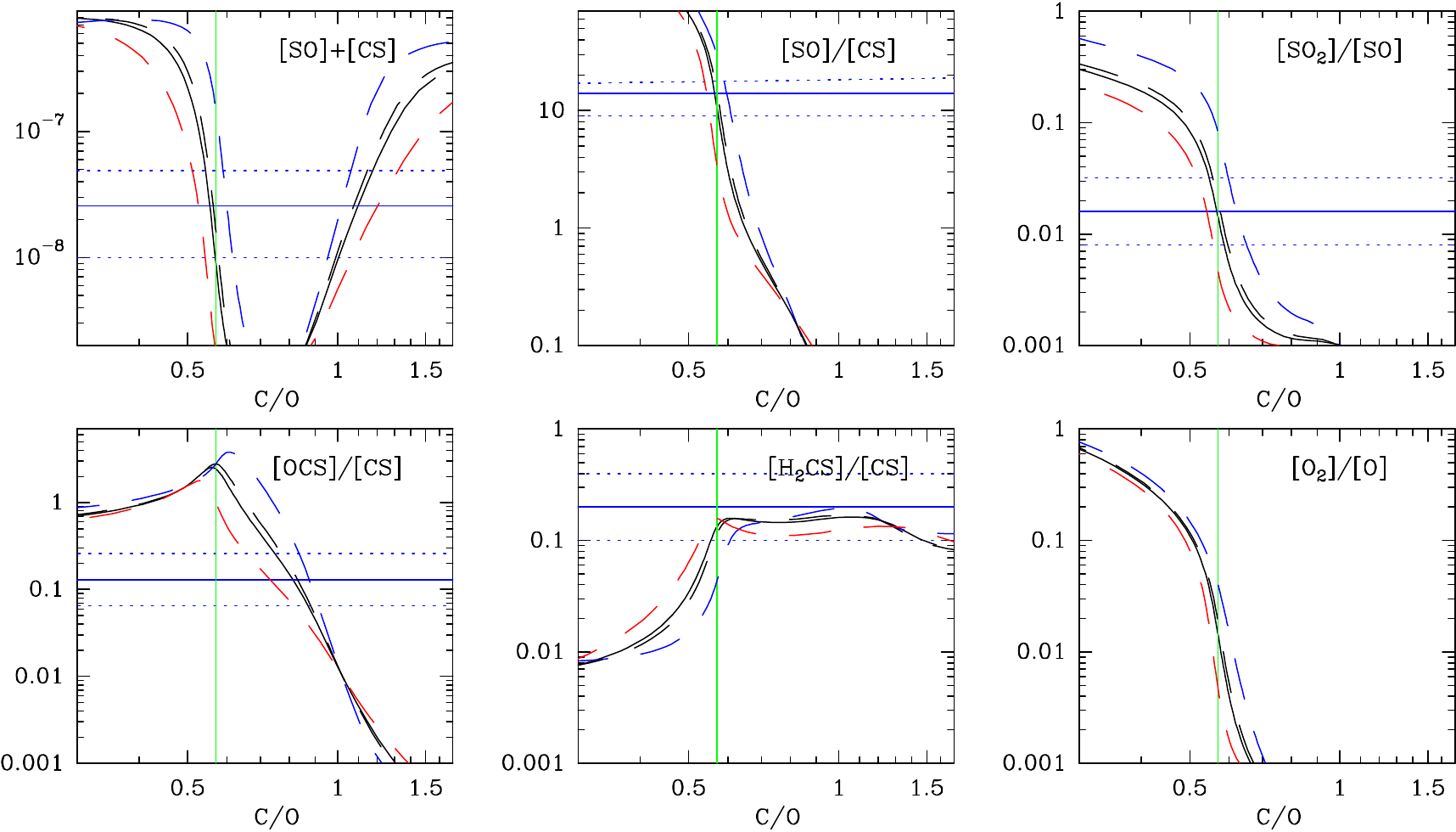}
   \caption{\label{fig12}
Steady-state calculations of fractional abundances and abundance ratios. All the input parameters
except O/H are the same as in Fig.~\ref{fig7}. Horizontal blue lines correspond to the 
observed values and their uncertainty. Vertical green lines indicate the estimated value of C/O elemental ratio,
C/O=0.6. The dashed blue, black, and red lines correspond to $\zeta_{H_2}$=1, 4 and 10$\times$10$^{-17}$~s$^{-1}$, 
respectively.
}
 \end{figure*}

In Table~\ref{tab_abundance} we show the fractional abundances of the studied species and some interesting
molecular abundance ratios. To estimate the fractional abundances for each velocity component, we 
assumed that the abundance of $^{13}$C$^{18}$O is the same for the two components and adopted a total 
molecular hydrogen column density of N(H$_2$)=7.6~10$^{22}$~cm$^{-2}$ \citep{Daniel13}.
Most of the fractional abundances and abundance ratios agree within the uncertainties for the two
velocity components. There are, however, significant chemical differences for
some species. One difference is that the HCO$^+$ deuterium fraction is a factor $\sim$3
higher in the 7~km~s$^{-1}$ component than in the 6.5~km~s$^{-1}$ one. The spectral velocity maps of
the DCO$^+$ 2$\rightarrow$1 and D$^{13}$CO$^+$ 2$\rightarrow$1 lines show that the 7~km~s$^{-1}$ component
is associated with the cold protostar B1b-N. 
The second difference is that the SO/CS abundance ratio is 2$-$10 times larger in the $\sim$6.5~km~s$^{-1}$ component than
in the $\sim$7.0~km~s$^{-1}$ component. High SO abundances, up to $\sim$10$^{-7}$, are found in bipolar outflows
because of the release of sulfuretted compounds into the gas phase \citep{Bachiller97}.
One could think that the high SO abundances could then result from the interaction of the cold core with 
the B1b-S outflow, or, more likely, of the interaction of the dense core with the B1a and/or B1d outflows.
However, the profiles of the SO lines are very narrow, which is contrary to one would expect if the emission were
coming from a shock or a turbulent medium. 
In Sect.~6, we discuss in detail the sulfur chemistry and possible explanations for the high SO abundance in Barnard 1.

One important result is that the [XH$^+$]/[X] ratios do not present variations between the
two velocity components. This supports our interpretation that the ionization fraction  
is mainly driven by CR and is consistent with the measured H$_2$ column density 
(7.6$\times$10$^{22}$~cm$^{-2}$). The core is expected to be thoroughly
permeable to CR and the ionization does not present significant fluctuations 
across the envelope.

\section{Chemical modeling}

We perform chemical modeling to constrain the physical and chemical properties
in Barnard B1b. We do not include here the accretion/desorption mechanisms of
gas-phase molecules on the grains and instead vary the amount of elemental
abundances of heavy atoms available in the gas phase to mimic these effects.
We explicitly consider, however, the formation on grains of the molecular hydrogen, H$_2$, and the 
isotopic substituted forms HD and D$_2$. Such an approach, although primitive,
allows us to disentangle the main gas-phase processes at work. We use both a
steady-state and time-dependent model to explore the chemistry and verified 
that the steady-state results are recovered by the time-dependent code
for sufficiently large evolution time (typically 10 Myr). 

Our chemical network contains 261 species linked through more than 5636 chemical reactions. Ortho and para forms 
of H$_2$, H$_2^+$, H$_3^+$, D$_2$, H$_2$D$^+$, D$_2$H$^+$, and D$_3^+$ are explicitly introduced, following the pioneering 
papers of \cite{Flower04,Flower06} for 
heavily depleted regions, and including the chemistry of carbon, oxygen, nitrogen, and sulfur. Discriminating between para and 
ortho forms of H$_2$ is particularly critical for deuteration in cold regions via the H$_3^+$ + HD $\leftrightarrow$ H$_2$D$^+$ + H$_2$ reaction as 
the endothermicity of the reverse reaction is of the same order of magnitude as the rotational energy of H$_2$, as first emphasized by \cite{Pagani92}.
Nitrogen chemistry is also very sensitive to the ortho and para forms of H$_2$, as pointed out by \citet{Lebourlot91}, because of the small endoergicity 
of the N$^+$ + H$_2$ $\rightarrow$ NH$^+$ + H  reaction. Whereas a definitive understanding of this reaction is still pending \citep{Zymak13}, we follow 
the prescription of \cite{Dislaire12} for the reaction rate coefficients of N$^+$ + H$_2$(p/o) reaction. 
Our chemical network also contains the recent updates coming from experimental studies of reactions involving neutral nitrogen atoms as 
reviewed in \cite{Wakelam13}. We also introduced species containing two nitrogen atoms, such as C$_2$N$_2$ and C$_2$N$_2$H$^+$, following the recent detection 
of C$_2$N$_2$H$^+$ by \cite{Agundez15} for which the reaction rate coefficients have essentially been taken from the UMIST astrochemical database \citep{Mcelroy13}.
An important feature that is not fully recognized is the link between sulfur and nitrogen chemistry as shown by the detection of NS, which is 
found both in 
regions of massive star formation and in cold dark clouds \citep{Irvine94,Irvine97}. The main formation reaction is through S + NH whose rate is given 
as 10$^{-10}$ cm$^3$ s$^{-1}$  in astrochemical databases, which then also induces that sulfur chemistry may depend on o/p ratio of H$_2$.

Our methodology is the following: We run the time-dependent model for a set of standard
elemental abundances and cosmic ray ionization rates to determine which molecular abundances and 
molecular abundance ratios reach the steady state in a few 0.1$-$1~Myr, which is the estimated age for
Barnard B1b. These molecular abundances and abundance ratios are expected to be in equilibrium 
and are used to explore the parameter space using steady-state calculations.
On basis of the steady-state calculations, we select the most plausible values 
and run the time-dependent model to confirm that our fit is correct. In all our 
calculations, we assume uniform physical conditions and fix n(H$_2$)=10$^5$~cm$^{-3}$ and T$_k$=12~K, 
according with previous works \citep{Lis10,Daniel13}. We iterate several times to get the best match
with the observations. The predictions of our best-fit model are compared with observations 
in Figs.~\ref{fig7} and ~\ref{fig8}. 

\subsection{Cosmic ray ionization rate}

The [HCO$^+$]/[CO] ratio has been extensively used to estimate the cosmic ray ionization rate. The results
of our time dependent model confirms that it is, very likely, the most reliable diagnostic since it reaches the
equilibrium in 0.1~Myr and its chemistry is well known.
Other [XH$^+$]/[X] ratios, such as [HCS$^+$]/[CS] and [HCNH$^+$]/[HCN], are also expected to be in equilibrium
at the typical ages of a molecular cloud. 
Figure~\ref{fig9} shows the variation of the [HCO$^+$]/[CO], [HCNH$^+$]/[HCN], and [HCS$^+$]/[CS] as a function of 
the cosmic ray ionization rate in steady-state calculations. Assuming an uncertainty of a factor of 2 in the 
derived abundance ratios,
values of $\zeta_{H_2}$ in the interval between 3$\times$10$^{-17}$~s$^{-1}$ to 1$\times$10$^{-16}$~s$^{-1}$
are consistent with the observed [HCO$^+$]/[CO] and [HCNH$^+$]/[HCN] ratios.
These values are at the upper end of the values inferred for cloud cores
\citep{Caselli98} and are higher than those of the typical dark cloud 
L1544, $\zeta_{H_2}$ $\sim$1$\times$10$^{-18}$ \citep{Galli15}. 
This could explain the higher gas and dust temperature in this dense core of T$_d$$\sim$12~K,
compared with dust temperature in L1544 of T$_d$$<$6$-8$~K \citep{Crapsi07}. 

The [HCS$^+$]/[CS] ratio is not fitted by any reasonable value of
$\zeta_{H_2}$. This is not surprising since the HCS$^+$ emission is mainly coming from the low density 
envelope. In Fig.~\ref{fig7} we show model predictions for a density of
n(H$_2$)=10$^4$~cm$^{-3}$. The abundances of most molecules decrease while  
the HCS$^+$ abundance remains constant and that of CCS increases. These calculations were carried out assuming the same elemental abundances as
in the high density case. This is not realistic since lower depletions are expected in
the envelope and, therefore, the HCS$^+$ abundance would be higher. Moreover, this 
core is illuminated on its eastern side by the nearby Herbig Ae stars LkH$\alpha$~327 and 
LkH$\alpha$~328  \citep{Walaw09,Young15}. The UV photons are expected to increase the
HCS$^+$ abundance. In fact, the HCS$^+$ 2$\rightarrow$1 emission
is maximum toward the illuminated side (see Fig.~\ref{fig2}). 

Our estimate of the cosmic ray ionization rate permits us to
obtain an estimate of the N$_2$ abundance in this core. 
Since it is one of the
main nitrogen reservoirs, N$_2$ cannot
be observed at (sub)millimeter wavelengths because of its lack of 
dipole moment. We can only estimate its abundance indirectly through the
protonated compound, N$_2$H$^+$.
The N$_2$H$^+$ abundance is very sensitive to the cosmic ray ionization rate.  
In Fig.~\ref{fig9}, we present the [N$_2$H$^+$]/[N$_2$] ratio as a function of
the cosmic ray ionization rate. Assuming $\zeta_{H_2}$=4$\times$10$^{-17}$~s$^{-1}$
and on the basis of our observational data, we derive [N$_2$]$\sim$8$\times$10$^{-5}$,
which is consistent with the typical value of the N elemental abundance in
a dark cloud, N/H=6.4$\times$10$^{-5}$.

The HOCO$^+$ molecule (protonated carbon dioxide) is thought to form via
a standard ion-molecule reaction, the transfer of a proton from H$_3^+$
to CO$_2$ (see, e.g., \citealp{Turner99}). To date HOCO$^+$ has been 
detected toward the Galactic center, a few
starless translucent clouds \citep{Turner99}, a single low-mass Class 0 protostar \citep{Sakai08}, and 
the protostellar shock L1157-B1 \citep{Podio14}. 
Interestingly, this molecule is particularly abundant, $\sim$10$^{-9}$,   
in the Galactic center region \citep{Minh88,Minh91,Deguchi06,Neill14}. 
We derive an HOCO$^+$ abundance of $\sim$10$^{-11}$, i.e, two orders of magnitude lower than
that derived in the Galactic center clouds \citep{Deguchi06,Neill14} and L1157-B1 \citep{Podio14}, 
and a factor of $\sim$10 lower than
that in translucent clouds \citep{Turner99}. This is not unexpected sine the HOCO$^+$ abundance is
inversely proportional to the gas density \citep{Sakai08} and B1b is a dense core. Our model 
underpredicts the HOCO$^+$ abundance by a factor 200 (see Fig.~\ref{fig7}). As commented below, 
HOCO$^+$ is proxy of CO$_2$, which is not efficiently formed in the gas phase \citep{Turner99}.

The lack of an electric dipole
moment of CO$_2$ makes gas-phase detections
difficult. The abundance of CO$_2$ in the solid phase has been investigated 
through its two vibrational modes at 4.3 and
15.0~$\mu$m. In the solid phase,
the typical abundance is $\sim$10$^{-6}$ relative to total H$_2$ \citep{Gerakines99,Gibb04}. In
the gas phase, the typical detected abundance is 
1$-$3$\times$10$^{-7}$ \citep{vandishoeck96,Boonman03, Sonnen06}.
Assuming the [HOCO$^+$]/[CO$_2$] ratio
predicted by our model, 2.2$\times$10$^{-4}$, we estimate a CO$_2$ abundance 
of $\sim$5$\times$10$^{-8}$. 
This value is at the lower end of those measured in star-forming regions 
but 200 times larger than that predicted by our gas-phase model. Photodesorption and 
sputtering of the icy mantles are more likely to be the main CO$_2$ production mechanisms.

\subsection{Sulfur chemistry: C, O, and S elemental abundances}

The initial elemental abundances in the gas phase are an essential parameter in the chemical
modeling. In cloud cores (A$_v$$>$10~mag), as long as the C elemental abundance is lower 
than O, we can assume that the abundance of 
C is equal to that of CO and conclude that  C/H=1.7$\times$10$^{-5}$;  we note that fractional abundances are given with respect to H$_2$. 
The main source of uncertainty in this value comes from the uncertainty in the CO abundance, 
which is estimated to be a factor of 2.

To estimate the amount of S in the gas phase requires a detailed modeling of its chemistry. 
As shown in  Fig.~\ref{fig10}, the cosmic ray ionization rate is a critical paramenter to
estimate the S/H. The main sulfur reservoirs in a dense core are atomic sulfur (S) and SO (see Fig.~\ref{fig11}). Unfortunately, chemical models make a poor work in predicting the fractional abundance of SO, 
because the predicted SO abunance depends on two uncertain factors: i) the 
S + O$_2$ $\rightarrow$ SO + O and S + OH $\rightarrow$ SO + H reaction rates and ii) the O$_2$ and OH abundances.

\begin{table}
\caption{Elemental abundances relative to H}
\label{tab_model}
\begin{tabular}{lll|ll|ll}\\ \hline \hline
\multicolumn{3}{c|}{Solar} & \multicolumn{2}{c|}{Dark cloud: TMC1$^1$} & \multicolumn{2}{c}{Barnard 1b$^2$} \\
\multicolumn{1}{c}{} & \multicolumn{1}{c}{X/H$^3$} & \multicolumn{1}{c|}{f$_D$$^4$} &
\multicolumn{1}{c}{X/H} & \multicolumn{1}{c|}{f$_D$} & \multicolumn{1}{c}{X/H} & \multicolumn{1}{c}{f$_D$} \\
 \hline
 C     &  1.4~10$^{-4}$ &  $\sim$1 & 1.4~10$^{-4}$   &  $\sim$1 &  1.7~10$^{-5}$ &   $\sim$8  \\  
 O     &  3.1~10$^{-4}$ &  $\sim$1 & 1.3~10$^{-4}$  &  $\sim$2  &  3.0~10$^{-5}$ &   $\sim$10  \\
 N     &  7.7~10$^{-5}$ &  $\sim$1 &  6.2~10$^{-5}$  &  $\sim$1  & 6.4~10$^{-5}$ &   $\sim$1  \\
 S     &  1.5~10$^{-5}$ &  $\sim$1 &   8.0~10$^{-8}$  &  $\sim$200 &   6.0~10$^{-7}$ &   $\sim$25 \\ \hline \hline
\end{tabular}

\noindent
$^1$ From the compilation by \citet{Agundez13}; 
$^2$ This work;\\  
$^3$ Elemental abundance in the gas phase; 
$^4$ Depletion factor = (X/H)$_{\odot}$/(X/H).      
\end{table}

The S + O$_2$ reaction rate coefficient at low temperatures is not known. 
Experimental measurements have only been performed above 200~K.
\citet{Lu04} reviewed the various theoretical and 
experimental results and adopted a value of 2 $\times$ 10$^{-12}$ cm$^3$ s$^{-1}$ at 298~K. 
The UMIST and KIDA databases suggest different values. In KIDA, a constant value of 
2.1 $\times$ 10$^{-12}$ cm$^3$ s$^{-1}$  is suggested in 
the 250$-$430~K temperature range whereas the latest version of the UMIST database 
gives 1.76 $\times$ 10$^{-12}$ (T/300)$^{0.81}$ exp(30.8/T) cm$^3$ s$^{-1}$ in the 200$-$3460~K temperature interval. 
This small activation barrier would imply rates as low as 10$^{-23}$~cm$^3$~s$^{-1}$  at the temperature of 12~K
and a decrease in the SO, SO$_2$, and SO$^+$ abundances. 
Quasiclassical trajectory calculations were carried out for the reaction S + O$_2$ and its reverse
by \citet{Rodrigues03}. With 6000 trajectories they obtain that the reaction rate remains constant or
decreases very slowly for temperatures between 50~K and 1000~K.
Using the same potential, we performed new quasiclassical
trajectory calculations with 100 000 trajectories for temperatures
between 10~K and 50~K (see Appendix A).
Our results confirm that the reaction rate remains very
constant (within 30\%) in the 10~K to 50~K range. 
Therefore, we adopted the KIDA value, 
2.1 $\times$ 10$^{-12}$ cm$^3$ s$^{-1}$, in our calculations.

The second more important reaction for the formation of SO is S + OH. \citet{DeMore97} measured a value of 
6.6 $\times$ 10$^{-11}$ cm$^3$~s$^{-1}$ at 398~K, which is the value adopted in our calculations. Recent theoretical calculations
by \citet{Goswami14} predicted a value a factor of $\sim$2 lower. These calculations were performed for a range of
temperatures between 10$-$500~K and the reaction rate remains very constant over this range with an increase of a factor 
of $\sim$1.6 for temperatures close to 10~K. Our adopted  experimental value is very likely good by a factor of 2 for
the physical conditions of B1.

The oxygen and sulfur chemistries are tightly linked. Since S is rapidly converted to SO by reacting with O$_2$ and OH, 
its abundance is very dependent on the amount of molecular oxygen in the gas phase, that itself depends on the C/O ratio 
(see Fig~\ref{fig12}). Conversely, a large abundance of S in the gas phase ($>$10$^{-6}$) leads
to a rapid destruction of O$_2$ and decreases the [O$_2$]/[O] ratio (see Fig~\ref{fig10}).
We find a good agreement with the observations, assuming C/O=0.6 and S/H
in the interval between 6$\times$10$^{-7}$ and 10$^{-6}$.
With these parameters we have an excellent fit of the SO abundance and the [SO]/[CS], [SO$_2$/[SO],
and [H$_2$CS]/[CS] ratios, although the agreement is worse for OCS (Fig~\ref{fig10} and~\ref{fig12}).  
We do not have a good explanation for the low observed abundance of OCS. It may be frozen onto
the grain mantles in this cold core. 
Taking into account the limitations of our model, of one single n(H$_2$) and T, we consider, however, that the
overall agreement with observational data is quite good.
Assuming these values, our chemical model predicts a steady-state O$_2$ abundance of 
$\sim$4$\times$10$^{-7}$. There is no measurement of the O$_2$ abundance in
B1b, but this predicted abundance is higher than the upper 
limits derived for other dark clouds \citep{Pagani03} and that reported in
$\rho$ Oph A ($\sim$5$\times$10$^{-8}$) \citep{Liseau12}.
The low O$_2$ abundance measured in dark clouds is usually explained 
in terms of time evolution. O$_2$ is a late-type molecule that presents low abundance 
($\sim$10$^{-8}$) at typical ages of dark clouds (a few 0.1 Myr). Another possibility is
that O$_2$ becomes frozen on grain mantles. Regarding the sulfur
chemistry, a lower O$_2$ abundance would decrease 
the SO production via the S + O$_2$ reaction. In this case, we would need to increase S/H
to reproduce the observations and our estimate would be a lower limit. 

Finally, our chemical model does not include surface chemistry. The main effect of dust
grains is to decrease the abundance of sulfur bearing molecules, especially 
SO$_2$, because of adsorption on grain mantles. The most abundant species in 
Barnard 1b,
SO, is thought to form in the gas phase. As long as this remains true, we consider that 
our measurement of the sulfur depletion is reliable.

\section{Barnard B1b chemistry}

We have used a state-of-the-art, gas-phase model to derive the cosmic ionization rate and elemental
abundances toward the dense core Barnard~1b. 
Our model fit predicts the observed abundances and abundance ratios reasonably well (within a factor of two), 
but it fails,  by several orders of magnitude to predict the abundances of HCS$^+$ and HOCO$^+$. 
As commented above, HCS$^+$ is very likely coming
from the lower density envelope instead of the dense core and cannot be accounted by our single 
point (n,T) model. 
In the case of HOCO$^+$, the model failure is not surprising since CO$_2$ is mainly formed
by surface reactions.

In Table~~\ref{tab_model} we compare the initial abundances derived toward Barnard~1b with typical values in 
dark clouds and diffuse clouds \citep{Agundez13}. The depletion of C and O is ten times larger
than those measured in prototypical dark clouds such as TMC~1 and L134N. These values are
similar to those found in prestellar cores and young stellar objects where high 
values of CO depletion ($\sim$5$-$10) and high deuterium fractions ($\sim$0.01) are measured 
\citep{Caselli99,Jorgensen02,Crapsi05,Empre09,Alonso10}. High values of the deuterium fractions 
are observed in Barnard 1b \citep{Lis02,Marcelino05}, which further confirm this similarity.
However, the relatively low S depletion in this source is unexpected.
A sulfur depletion factor of $\sim$100 is adopted to explain the
chemistry in dark clouds and dense cores \citep{Tafalla06,Agundez13}. 
A higher gas-phase sulfur abundance approaching the solar value of 1.5$\times$10$^{-5}$ has
only been found in bipolar 
outflows \citep{Bachiller97,Anderson13}, photodissociation regions \citep{Goicoechea06}, and 
hot cores \citep{Esplugues13,Esplugues14}. In these cases, this abundance was intrepreted as the consequence of 
the release of the S-species from the icy grain mantles because of thermal and nonthermal desorption 
and sputtering. Contrary to dark clouds and prestellar cores, in Barnard 1b the  S depletion is comparable
to that of C and O, which
explains the high abundances of S-bearing species in this core.

The abundance of sulfur in grains is very uncertain. Thus far,
the only S-bearing molecule unambiguously detected in ice
mantles is OCS, because of its large band strength in the infrared \citep{Geballe85,Palumbo95}, and 
maybe SO$_2$ \citep{Boogert97}, but there are only upper limits of the solid H$_2$S
abundance \citep{Jimenez11}, which is thought to be the main sulfur reservoir in the ice.
Measurements of the sulfur depletion are derived by comparing the predictions of 
gas-phase chemical models with the observed abundances of the main S-bearing species. 
As reported above, chemical models are limited by the uncertainty in
the reaction rates and the poorly known [O]/[O$_2$] ratio and the exact value of the 
sulfur depletion is uncertain. Our results suggest, however,
that the sulfur depletion in Barnard~1b is 
not as extreme as in dark clouds where the abundance of S-bearing species is lower.
 
This moderate sulfur depletion could be the consequence of the star formation activity in the region. 
B1b is irradiated on its eastern side by UV photons from the nearby Herbig Ae stars LkH$\alpha$~327 and 
LkH$\alpha$~328  \citep{Walaw09,Young15}. On its western 
side, B1b is impacted by the outflows associated with B1a and B1d \citep{Hatchell07}.
Not unrelated, the level of turbulence in B1 ($\sim$0.4 km s$^{-1}$) is higher than in cold 
cores such as L1544, TMC-1 or L134N  ($\sim$0.1 km s$^{-1}$). Sputtering induced by collisions may 
be efficient in this turbulent environment to erode the icy mantles and release the sulfuretted species to 
the gas phase. Sputtering could also explain the high abundances of complex molecules observed
in this core \citep{Marcelino09,Cernicharo12}. The main drawback of this interpretation is that it would also imply 
a low depletion of CO that is not observed.

Another possibility is that the moderate sulfur depletion is a consequence
of the rapid collapse of B1b. \citet{Aikawa05} investigated the impact of different 
collapse timescales on the gas chemistry. The initial conditions of their first model lie close to the 
critical Bonnor-Ebert sphere with a central density of $\sim$10$^4$~cm$^{-3}$.
When the central density of the core reaches 10$^5$~cm$^{-3}$, the S-bearing species were totally 
depleted at R$\sim$5000~AU. In the second case, internal gravity overwhelms pressure and the collapse
is much faster (about 0.1 Myr). When the central density of the core reaches 10$^5$~cm$^{-3}$, molecules 
such as CS, SO, and C$_2$S have abundances of $\sim$10$^{-10}$ in the core center while CO is significantly
depleted. These abundances are too low compared with our measurements, but this is partially due to the 
adopted S elemental abundance in the Aikawa model (S/H=9.14$\times$10$^{-8}$). 
We have not targeted the position of the protostar but a nearby position
at R$\sim$2500~AU, where the sulfur depletions is expected to be lower. 
Moreover, our position is spatially coincident with the red wing of the B1b-S outflow.
Higher spatial resolution observations are needed to determine the possible influence of the B1b-S outflow on the S chemistry.

Summarizing, we propose that the low sulfur depletion in this region could be the result of two different
factors, which are both related to the star formation activity. The enhanced UV fields and the surrounding
outflows could be the cause of  peculiar initial conditions with low sulfur depletion and high abundances of complex
molecules. The star formation activity could also have induced a rapid collapse of the B1-b core that
preserves the high abundances of the sulfuretted species. The compact outflow associated with B1b-S
could also heat the surroundings and contribute to halting the depletion of S-molecules.

\section{Summary and conclusions}

On the basis of our spectral survey toward Barnard 1b, we selected a set of neutral and ionic species 
to determine the value of the cosmic ray ionization rate and depletion factors of the C, N, O, and S. 
These are the parameters that determine the gas ionization fraction and, hence, the dynamical evolution of this
core. In our survey, we detected the following ions:
HCO$^+$, H$^{13}$CO$^+$, HC$^{18}$O$^+$, HC$^{17}$O$^+$, DCO$^+$, D$^{13}$CO$^+$,
HOCO$^+$, DOCO$^+$, HCS$^+$, SO$^+$, HC$^{34}$S$^+$, DCS$^+$,
HCNH$^+$, H$_2$COH$^+$, HC$_3$NH$^+$,
NNH$^+$, NND$^+$, N$^{15}$NH$^+$, and $^{15}$NNH$^+$. This is the most complete inventory
of molecular ions in this core and, probably, the first secure detection of the deuterated ion 
DOCO$^+$ and DCS$^+$in any source.

We use a state-of-art, pseudo-time-dependent, gas-phase chemical model that includes the ortho and para forms of
H$_2$, H$_2^+$, D$_2^+$, H$_3^+$, H$_2$D$^+$, D$_2$H$^+$, D$_2$, and D$_3^+$. Our model 
assumes n(H$_2$)=10$^5$~cm$^{-3}$ and T$_k$=12~K, which is derived from previous works.
The observational data are well fitted with $\zeta_{H_2}$=(3$-$10)~$\times$10$^{-17}$~s$^{-1}$ and the elemental abundances 
O/H=3$\times$10$^{-5}$, N/H=(6.4$-$8)$\times$10$^{-5}$, C/H=1.7$\times$10$^{-5}$, and S/H=(6.0$-$10)~10$^{-7}$.
On basis of the
HOCO$^+$ and N$_2$H$^+$ observations, we derive abundances of 
$\sim$5$\times$10$^{-8}$ and (6.4$-$8)$\times$ 10$^{-5}$ for
CO$_2$ and N$_2$, respectively. 

Barnard 1b presents similar depletion of C and O as those measured in prestellar cores. The depletion of S is, however, moderate 
more similar to that found in bipolar outflows, hot cores, and 
   photon-dominated regions. This high C and O depletion, together with the moderate S depletion, produces
   a peculiar chemistry that is very rich in sulfuretted species.
   We propose that the low S depletion is the consequence of the peculiar initial conditions 
  (important outflows and enhanced UV fields in the surroundings) and a rapid collapse ($\sim$0.1 Myr) that 
   allows most S- and N-bearing species to remain in the gas phase. The compact outflow associated with B1b-S
 could also contribute to enhance the abundance of S-molecules.


\begin{acknowledgements}

We thank the Spanish MINECO for funding support from
 grants CSD2009-00038, FIS2012-32096, FIS2014-52172-C2 
AYA2012-32032, and ERC under ERC-2013-SyG, G. A. 610256 NANOCOSMOS.
We thank Prof. J.A.C. Varandas for sending us the Fortran code
of their potential energy surface for the S+O$_2$ reaction.
ER and MG thank the INSU/CNRS program PCMI for funding.
This research used the facilities of the Canadian Astronomy Data Centre operated by the National 
Research Council of Canada with the support of the Canadian Space Agency.

\end{acknowledgements}

\bibliography{B1}

\appendix

\section{Appendix}

The S+O$_2$ $\rightarrow$ SO +O
reaction  was studied using a quasiclassical
method with the most up-to-date
 potential energy surface for the ground singlet state
 of \citet{Rodrigues02}.
This reaction is exothermic by 
2345~K and presents a SO$_2$ deep well with a depth of 10831~K,
and there is no barrier along the minimum energy path when S approaches O$_2$.
This potential, however, presents several minima connected
among them with saddle points.

\begin{figure}[h]
\hspace*{-0.5cm}
\includegraphics[scale=0.40,angle=0]{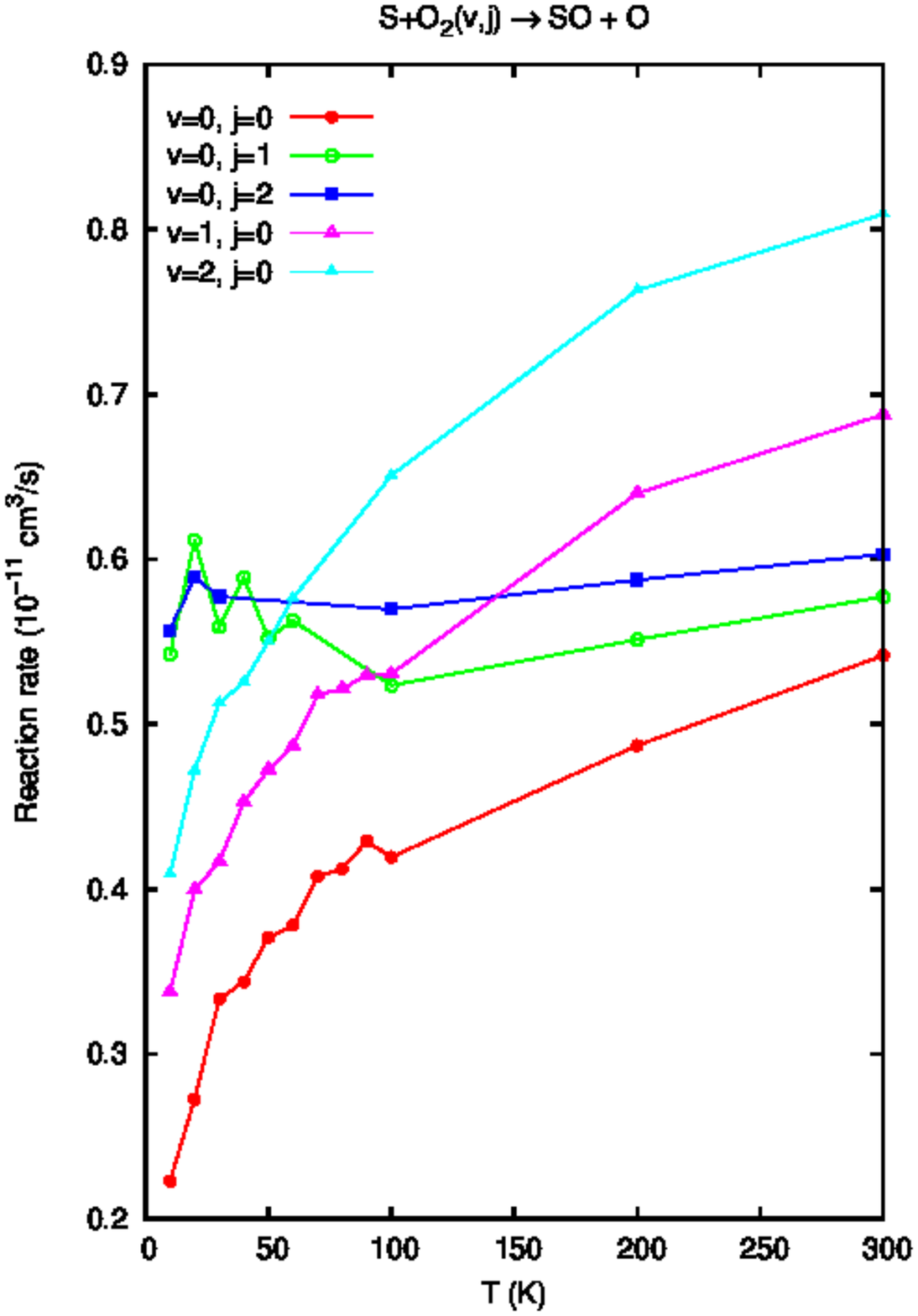}
   \caption{\label{rates-fig}
State-selected reaction rates for O$_2(v,j)$ + S $\rightarrow$ SO +O collisions
as a function of the translational temperatures for selected initial rovibrational
states of O$_2$.
}
\end{figure}

A quasiclassical method was employed to calculate
the rates for selected states of O$_2(v,j)$ and 
for temperatures in the 10-300~K temperature interval. This method is 
very analogous to that employed by \citet{Rodrigues03}
for higher temperatures. The MIQCT code was used \citep{Dorta15}
in which the initial conditions for a particular rovibrational state of O$_2$
are determined according to Karpus and coworkers \citep{Karplus65}.
The $E_{vj}$ energies of O$_2$ are determined by solving numerically
the monodimensional quantum Schr\"odinger equation. The initial translational
energy between the reagents is sampled from a Maxwell-Boltzmann distribution.
A set of $N_t$=10$^5$  initial
conditions are propagated for each initial rovibrational state and
temperature considered. The state-selected rate coefficients are then
obtained as
\begin{eqnarray}
k_{vj}(T) = g_e(T) \,\sqrt{ {8k_B T\over \pi \mu} }\, {N_r\over N_t}\, \pi b_{max}^2,
\end{eqnarray}
where $N_r$ denotes the number of reactive trajectories and $b_{max}$ is the maximum
impact parameter for which reaction takes place. Finally, $g_e$ is the electronic
partition function obtained assuming that only the ground singlet state,
 of the 27 spin-orbit states of the S($^3P$)+ O$_2(^3\Sigma_g^-)$,
is reactive, taking the form \citep{Rodrigues03}
\begin{eqnarray}
g_e(T)= \left\lbrace
          3\left( 5+3 e^{-570/T} + e^{-825/T}\right)
        \right\rbrace^{-1}.
\end{eqnarray}
The results obtained are shown in Fig.~\ref{rates-fig}. 

For the initial rotational state, $j=0$, the rate increases with
temperature, while for $j=$1 or 2 the rate is nearly
constant. This behavior is attributed
to  the series of wells and saddle points in the entrance channel: when O$_2$ is not rotationally excited and finds one of these barriers is reflected back.
On the contrary, when O$_2$
is rotationally excited it overcomes those barriers
leading to reaction. This is not a simple question of excess of 
energy, since the rate also increases with temperature for $v=$1 or 2 and $j=0$.
 
At 10~K, the population in $j$=0 is less than 20\%, and the average reaction
rate is 4.8$\times$$10^{-12}$~cm$^3$~s$^{-1}$, which is higher than the value in the KIDA datebase
of 2.1$\times$$10^{-12}$~cm$^3$~s$^{-1}$,  and can be considered approximately constant since it only 
varies between 4.8 and 5.5 for temperatures between 10 and 50~K.

\onecolumn
\section{Tables}
\begin{longtable}{lllllll} 
\caption{Line integrated intensities}\\ \hline \hline
\multicolumn{1}{c}{Mol.}    &  \multicolumn{1}{c}{Trans.} &   \multicolumn{1}{c}{Freq.}     &  
\multicolumn{1}{c}{E$_u$}   &  \multicolumn{1}{c}{6.5~km~s$^{-1}$}    & \multicolumn{1}{c}{7.0~km~s$^{-1}$}   & \\
 &  & \multicolumn{1}{c}{(MHz)}     &  \multicolumn{1}{c}{(K)}   & \multicolumn{1}{c}{Area(K~km~s$^{-1}$)}  & 
\multicolumn{1}{c}{Area(K~km~s$^{-1}$)} &  \\ \hline
C$^{18}$O     &    1$\rightarrow$0 & 109782.16  &  5.3    &  1.520 (0.001)    &   3.559 (0.001)  & \\
              &    2$\rightarrow$1              &  219560.36   &  15.8 & 0.568 (0.008)  &  3.088 (0.010) & \\ 
$^{13}$C$^{18}$O & 1$\rightarrow$0 & 104711.39  &  5.0    &  0.029 (0.002)    &   0.054 (0.003) &   \\
                 & 2$\rightarrow$1 & 209419.16  &  15.1   & 0.019 (0.006)     &   0.085 (0.007)  &\\
C$^{17}$O     &    1$\rightarrow$0 & 112359.29  &  5.4    &  0.091 (0.004)    &   0.756 (0.005) &  \\

H$^{13}$CO$^+$ &   1$\rightarrow$0 &  86754.29  &  4.2    &  1.545 (0.002)    &   1.553 (0.003)  & \\
               &   3$\rightarrow$2 & 260255.34  &  25.0   & 0.329 (0.005)     &  0.323 (0.006) & \\
HC$^{18}$O$^+$ &   1$\rightarrow$0 &  85162.22  &  4.1    &  0.209 (0.002)    &   0.148 (0.003) &  \\
               &  3$\rightarrow$2  & 255479.39  &  24.5  & 0.042 (0.005)  &  0.010 (0.006)  & \\
HC$^{17}$O$^+$ &   1$\rightarrow$0 &  87057.53  &  4.2    & 0.012 (0.003)  &  0.028 (0.003) & \\
DCO$^+$        &  2$\rightarrow$1   & 144077.24 & 10.4  & 1.943 (0.248)  &  2.579 (0.248)  &  Component at $\sim$6.0 km~s$^{-1}$ \\
               &  3$\rightarrow$2   & 216112.58  & 20.7  & 1.535 (0.004)  &  2.008 (0.004) &  \\
D$^{13}$CO$^+$ &  2$\rightarrow$1   & 141465.13  & 10.2  &  0.169 (0.006) &  0.342 (0.008) &  \\
               &  3$\rightarrow$2   & 212194.49  & 20.4  & 0.103 (0.003)  &  0.080 (0.006) &  \\
HOCO$^+$   &   4$_{0,4}$$\rightarrow$3$_{0,3}$ &  85531.51  &  10.3   & 0.082 (0.002)  &  0.043 (0.002) &  \\
           &   5$_{0,5}$$\rightarrow$4$_{0,4}$ & 106913.56  &  15.4   & 0.077 (0.002)  &  0.025 (0.003) &  \\
           &   10$_{0,10}$$\rightarrow$9$_{0,9}$  & 213813.385 &  56.4   & 0.010 (0.003)  &  0.001 (0.002) &  \\
DOCO$^+$   &   5$_{0,5}$$\rightarrow$4$_{0,4}$ & 100359.55  &  14.5  & 0.008 (0.001)  &  0.005 (0.002) & \\
SO         &  2$_2$$\rightarrow$1$_1$  &   86093.96 & 19.3  & 1.322 (0.001)  &  0.573 (0.001) & \\
           &  2$_3$$\rightarrow$1$_2$  &   99299.89 &  9.2  & 3.208 (0.001)  &  2.410 (0.002) & \\
           &  5$_4$$\rightarrow$4$_4$  &  100029.55 & 38.6  & 0.048 (0.002)  &  0.004 (0.002) & \\
           &  3$_2$$\rightarrow$2$_1$  &  109252.18 & 21.1  & 1.393 (0.003)  &  0.533 (0.004) & \\
           &  3$_4$$\rightarrow$2$_3$  &  138178.65 & 15.9  &  2.021 (0.005) &  2.129 (0.005) & \\
           &  4$_3$$\rightarrow$3$_2$  &  158971.81 & 28.7  & 0.928 (0.002)  &  1.009 (0.003) & \\
           &  5$_4$$\rightarrow$4$_3$  &  206176.01 & 38.6  & 0.717 (0.006)  &  0.240 (0.007) & \\
           &  5$_5$$\rightarrow$4$_4$  &  215220.65 & 44.1  & 0.613 (0.007)  &  0.205 (0.008) & \\
           &  5$_6$$\rightarrow$4$_5$  &  219949.39 & 35.0  & 1.271 (0.007)  &  0.573 (0.009) & \\
           &  2$_1$$\rightarrow$1$_2$  &  236452.29 & 15.8  & 0.098 (0.009)  &  0.033 (0.010) & \\
           &  3$_2$$\rightarrow$2$_3$  &  246404.59 & 21.1  & 0.040 (0.006)  &  0.035 (0.007) & \\
           &  6$_7$$\rightarrow$5$_6$  &  261843.70 & 47.6  & 0.813 (0.008)  &  0.312 (0.010) & \\
           &  4$_3$$\rightarrow$3$_4$  &  267197.74 & 28.7  & 0.010 (0.007)  &  0.025 (0.008) & \\

$^{34}$SO  &  2$_2$$\rightarrow$1$_1$  &   84410.68 &  19.2  & 0.071 (0.002)  &  0.020 (0.002) & \\
           &  2$_3$$\rightarrow$1$_2$  &   97715.40 &   9.1  & 0.748 (0.002)  &  0.349 (0.002) & \\
           &  3$_2$$\rightarrow$2$_1$  &  106743.36 &  20.9  & 0.090 (0.002)  &  0.018 (0.002) & \\
           &  3$_4$$\rightarrow$2$_3$  &  135775.65 &  15.6  &  0.578 (0.003) &  0.170 (0.004) & \\
           &  4$_3$$\rightarrow$3$_2$  &  155506.80 &  28.4  &  0.071 (0.006) &  0.062 (0.008) & \\
           &  4$_4$$\rightarrow$3$_3$  &  168815.11 &  33.4  &  0.069 (0.008) &  0.018 (0.009) & \\
           &  5$_4$$\rightarrow$4$_3$  &  201846.65 &  38.1  &  0.047 (0.008) &  0.001 (0.009) & \\
           &  5$_5$$\rightarrow$4$_4$  &  211013.02 &  43.5  &  0.049 (0.005) &  0.001 (0.005) & \\
           &  5$_6$$\rightarrow$4$_5$  &  215839.92 &  34.4  & 0.142 (0.005)  &  0.060 (0.006) & \\ 
           &  6$_5$$\rightarrow$5$_4$  &  246663.39 &  49.9  & 0.017 (0.005)  &  0.004 (0.006) & \\
           &  6$_6$$\rightarrow$5$_5$  &  253207.02 &  55.7  & 0.001 (0.006)  &  0.020 (0.004) & \\ 
           &  6$_7$$\rightarrow$5$_6$  &  256877.81 &  46.7  & 0.044 (0.005)  &  0.016 (0.006) & \\
$^{33}$SO  &  2 3 7/2$\rightarrow$1 2 7/2 &   98443.84 &  9.2  & 0.006 (0.001)  &  0.001 (0.002) & \\
           &  2 3 5/2$\rightarrow$1 2 5/2 &   98455.19 &  9.2  & 0.008 (0.001)  &  0.001 (0.002) & \\
           &  2 3 3/2$\rightarrow$1 2 3/2 &   98460.49 &  9.2  & 0.004 (0.002)  &  0.007 (0.002) & \\    
           &  2 3 3/2$\rightarrow$1 2 1/2 &   98474.60 &  9.2  & 0.024 (0.001)  &  0.001 (0.002) & \\ 
           &  2 3 5/2$\rightarrow$1 2 3/2 &   98482.30 &  9.2  & 0.039 (0.002)  &  0.001 (0.002) & \\
           &  2 3 7/2$\rightarrow$1 2 5/2 &   98489.23 &  9.2  & 0.047 (0.001)  &  0.011 (0.004) & \\
           &  2 3 9/2$\rightarrow$1 2 7/2 &   98493.64 &  9.2  & 0.074 (0.001)  &  0.011 (0.004) & \\
           &  3 2 7/2$\rightarrow$2 1 5/2 &  107953.80 & 21.0   & 0.006 (0.002)  &  0.001 (0.007)  & \\
           &  3 4 7/2$\rightarrow$2 3 5/2  & 136939.36  & 15.7 & 0.035 (0.004) &  0.007 (0.005) & \\
           &  3 4 9/2$\rightarrow$2 3 7/2  & 136943.67  & 15.7 & 0.014 (0.014) &  0.039 (0.005) & Component at$\sim$ 6.0 km~s$^{-1}$ \\
              & 3 4 11/2$\rightarrow$2 3 9/2 &   136946.19 & 15.7  & 0.041 (0.004) &  0.025 (0.005) & \\
              & 4 3 7/2$\rightarrow$2 3 5/2  &   157173.54 &  28.5 & 0.023 (0.004) &  0.001 (0.005) & \\
              & 4 3 5/2$\rightarrow$2 3 3/2  &   157173.54 &  28.5 &               &                & \\
S$^{18}$O  &  2$_3$$\rightarrow$1$_2$  &   93267.38  &  8.7 & 0.162 (0.002)  &  0.052 (0.002) & \\
           &  3$_2$$\rightarrow$2$_1$  &   99803.66  & 20.5 & 0.012 (0.001)  &  0.001 (0.004) & \\
           &  4$_3$$\rightarrow$3$_2$  &   145874.49 &  27.5 & 0.015 (0.004) &  0.008 (0.005) & \\ 
           &  4$_4$$\rightarrow$3$_3$  &   159428.31 &  32.4 & 0.015 (0.005) &  0.001 (0.005) & \\
           &  4$_5$$\rightarrow$3$_4$  &  166285.31  &  22.9 & 0.061 (0.005) &  0.058 (0.007) & \\
           &  5$_6$$\rightarrow$4$_5$  &  204387.94  &  32.7 & 0.027 (0.008)  &  0.012 (0.009) &  \\
SO$^+$     &  5/2$\rightarrow$3/2 (e)  & 115804.40 &  8.9  & 0.157 (0.008)  &  0.111 (0.010) &  \\
           &  5/2$\rightarrow$3/2 (f)  & 116179.95 &  8.9  & 0.196 (0.007)  &  0.071 (0.008) &  \\
           &  7/2$\rightarrow$5/2 (e)  & 162198.60 & 16.7  & 0.075 (0.003) &  0.115 (0.004)  &  \\
           &  7/2$\rightarrow$5/2 (f)  & 162574.06 & 16.7  & 0.084 (0.004) &  0.117 (0.004)  &  \\
           &  9/2$\rightarrow$7/2 (e)  & 208590.02 & 26.7  & 0.079 (0.006)  &  0.008 (0.007) &  \\
           & 9/2$\rightarrow$7/2 (f)   & 208965.42 & 26.8  & 0.067 (0.005)  &  0.020 (0.006) &  \\
           & 11/2$\rightarrow$9/2 (e)  & 254977.93 &  38.9  & 0.025 (0.005)  &  0.019 (0.006) &  \\
           & 11/2$\rightarrow$9/2 (f)  & 255353.24 &  39.0  & 0.026 (0.005)  &  0.019 (0.006) & \\
CS         & 2$\rightarrow$1 &  97980.95  & 7.1  &  2.250 (0.038)  &  1.064 (0.001) &  Self-absorbed \\
           & 3$\rightarrow$2 & 146969.02  & 14.1 &  1.301 (0.036)  &  1.483 (0.043) &  Self-absorbed\\
           & 5$\rightarrow$4 & 244935.55  & 35.3 &  0.676 (0.021)  &  0.455 (0.025) &  Bad fit \\
C$^{34}$S  & 2$\rightarrow$1          & 96412.95 & 6.9 &   0.578 (0.003) &  0.505 (0.004) & \\
           & 3$\rightarrow$2          & 144617.10 & 13.9 & 0.298 (0.006) & 0.213 (0.006)  & \\ 
           & 5$\rightarrow$4          & 241016.09 & 34.7 & 0.038 (0.006) & 0.021 (0.007)  &  \\
$^{13}$CS        & 2$\rightarrow$1    & 92494.27 & 6.7 &  0.259 (0.004) &  0.159 (0.005) &  \\
                 & 3$\rightarrow$2    & 138739.26 & 13.3 & 0.070 (0.004) & 0.162 (0.005) &  \\ 
                 & 5$\rightarrow$4    & 231220.68 & 33.3 & 0.032 (0.006) & 0.001 (0.027) &  \\
$^{13}$C$^{34}$S & 2$\rightarrow$1    & 90925.99 & 6.5 & 0.009 (0.003) &  0.007 (0.003) &  \\ 

C$^{33}$S        & 2$\rightarrow$1    & 97172.06  &  7.0 & 0.103 (0.006) &  0.085 (0.007) & \\
                 & 3$\rightarrow$2    & 145755.73 & 14.0 & 0.044 (0.006) & 0.037 (0.007) &  Bad fit \\ 

HCS$^+$  & 2$\rightarrow$1     & 85347.87  &  6.1  & 0.122 (0.002)  &  0.110 (0.002)  &  \\
         & 4$\rightarrow$3     & 170691.62 & 20.5   & 0.014 (0.025)  &  0.137 (0.029) &  \\
         & 5$\rightarrow$4     &  213360.65 & 30.7  & 0.037 (0.004)  &  0.048 (0.005) &  \\
         & 6$\rightarrow$5     &  256027.11 & 43.0  & 0.016 (0.004)  &  0.001 (0.005) &  \\
HC$^{34}$S$^+$ &  2$\rightarrow$1                 & 83965.63  &  6.0  & 0.017 (0.002)  &  0.001 (0.002) &  \\
DCS$^+$    &  3$\rightarrow$2                 & 108108.01 & 10.4  & 0.008 (0.002)  &  0.001 (0.002) &  \\
HCNH$^+$     & 2$\rightarrow$1       & 148221.46 & 10.7   & 0.032 (0.003)  &  0.032 (0.004) &  \\
             & 3$\rightarrow$2          &  222329.30 & 21.3  & 0.050 (0.007)  &  0.022 (0.008) &  \\
o$-$H$_2$$^{13}$CO  &  2$_{12}$$\rightarrow$1$_{11}$  &  137449.95 & 6.6 &  0.135 (0.007)   &  0.149 (0.008) &  \\
                    &  2$_{11}$$\rightarrow$1$_{10}$  &  146635.67 & 7.3 &  0.118 (0.005)   &  0.064 (0.006) & Red wing \\
                    &  3$_{13}$$\rightarrow$2$_{12}$  &  206131.62 & 16.5 & 0.074 (0.006)   &  0.041 (0.007) &   \\
                    &  3$_{12}$$\rightarrow$2$_{11}$  &  219908.48 & 17.8 & 0.040 (0.005)   &  0.033 (0.005) &   \\ 
p$-$H$_2$$^{13}$CO  &  2$_{02}$$\rightarrow$1$_{01}$  &  141983.75 & 10.2 &  0.079 (0.006)   &  0.072 (0.007) &   \\
                    &  3$_{03}$$\rightarrow$2$_{02}$  &  212811.19 & 20.4 &  0.053 (0.007)   &  0.016 (0.009) &   \\
H$_2$COH$^+$ & 1$_{1,0}$$\rightarrow$1$_{0,1}$  &  168401.14 & 11.1 & 0.012 (0.006)  &  0.019 (0.008) &  \\
             & 1$_{1,1}$$\rightarrow$0$_{0,0}$  &  226746.31 & 10.9  & 0.001 (0.004)  &  0.021 (0.008) &  \\  
             & 4$_{0,4}$$\rightarrow$3$_{0,3}$  &  252870.34 & 30.4  & 0.001 (0.006)  &  0.018 (0.006) &  \\ 
OCS          & 7$\rightarrow$6  & 85139.10   & 16.3  & 0.174 (0.002) & 0.007 (0.003) &  \\
             & 8$\rightarrow$7  & 97301.21   & 21.0  & 0.167 (0.002) & 0.058 (0.003) &  \\
             & 9$\rightarrow$8  & 109463.06  & 26.3  & 0.130 (0.002) & 0.055 (0.003) & \\
             & 11$\rightarrow$10  & 133785.90 & 38.5 & 0.055 (0.004) & 0.055 (0.005) & Red wing \\
             & 12$\rightarrow$11  & 145946.81 & 45.5 & 0.034 (0.004) & 0.043 (0.005) &   \\
             & 13$\rightarrow$12  & 158107.36 & 53.1 & 0.042 (0.005) & 0.029 (0.006) & \\
OC$^{34}$S   & 7$\rightarrow$6     &  83057.97   &  15.9  & 0.011 (0.003)  &              &   \\
             & 8$\rightarrow$7     &  94922.80   &  20.5  & 0.009 (0.002)  &              &   \\
             & 9$\rightarrow$8     &  106787.39  &  25.6  & 0.006 (0.003)  &              &   \\  
H$_2$$^{13}$CS  & 3$_{1,3}$$\rightarrow$2$_{1,2}$ &  97632.20   &  7.8  & 0.010 (0.002)  &     &   \\
                & 3$_{1,2}$$\rightarrow$2$_{1,1}$ &  100534.75  &  8.0  & 0.007 (0.001)  &     &   \\
                & 3$_{0,3}$$\rightarrow$2$_{0,2}$ &   99077.84  &  9.5  & 0.006 (0.002)  &     &   \\
SO$_2$          & 8$_{1,7}$$\rightarrow$8$_{0,8}$ &   83688.09 &  36.7  & 0.079 (0.003)  & 0.028 (0.004) &  \\
                & 2$_{2,0}$$\rightarrow$3$_{1,3}$ &  100878.11 &  12.6  & 0.013 (0.002)  &               &   \\             
                & 3$_{1,3}$$\rightarrow$2$_{0,2}$ &  104029.43 &   7.7  & 0.677 (0.002)  & 0.216 (0.002) &  \\ 
                & 10$_{1,9}$$\rightarrow$10$_{0,10}$ &  104239.30 &  54.7 & 0.020 (0.002)  &       &       \\
                &  8$_{0,8}$$\rightarrow$7$_{1,7}$   &  116980.45 &  32.7 & 0.197 (0.015)  &       &    \\
                &  5$_{1,5}$$\rightarrow$4$_{0,4}$   &  135696.02 &  15.7 & 0.453 (0.005)  & 0.133 (0.005) &  \\
                &  6$_{2,4}$$\rightarrow$6$_{1,5}$   &  140306.17 &  29.2 & 0.077 (0.004)  & 0.026 (0.005) &   Red wing \\ 
                &  4$_{2,2}$$\rightarrow$4$_{1,3}$   &  146605.52 &  19.0 & 0.168 (0.005)  &            &  \\
                &  2$_{2,0}$$\rightarrow$2$_{1,1}$   &  151378.66 &  12.6 & 0.131 (0.005)  & 0.052 (0.006) &  \\
                &  3$_{2,2}$$\rightarrow$3$_{1,3}$   &  158199.78 &  15.3 & 0.222 (0.005)  &        &    \\
                &  10$_{0,10}$$\rightarrow$9$_{1,9}$ &  160827.84 &  49.7 & 0.060 (0.007)  &  &  \\
                &  5$_{2,4}$$\rightarrow$5$_{1,5}$   &  165144.65 &  23.6 & 0.127 (0.006)  &  &  \\
                &  7$_{1,7}$$\rightarrow$6$_{0,6}$   &  165225.45 &  27.1 & 0.239 (0.006)  &  &  \\ 
                &  3$_{2,2}$$\rightarrow$2$_{1,1}$   &  208700.34 &  15.3 & 0.237 (0.006)  & 0.095 (0.007) & \\
                &  4$_{2,2}$$\rightarrow$3$_{1,3}$   &  235151.72 &  19.0 & 0.242 (0.007)  & 0.042 (0.008) & \\
                &  5$_{2,4}$$\rightarrow$4$_{1,3}$   &  241615.80 &  23.6 & 0.154 (0.006)  & 0.042 (0.007) & \\
                &  6$_{3,3}$$\rightarrow$6$_{2,4}$   &  254280.54 &  41.4 & 0.019 (0.005)  &       &  \\
                &  4$_{3,1}$$\rightarrow$4$_{2,2}$   &  255553.30 &  31.3 & 0.020 (0.005)  &       &  \\
                &  3$_{3,1}$$\rightarrow$3$_{2,2}$   &  255958.04 &  27.6 & 0.029 (0.005)  &       &   \\
                &  5$_{3,3}$$\rightarrow$5$_{2,4}$   &  256246.95 &  35.9 & 0.021 (0.005)  &       &   \\
                &  7$_{2,6}$$\rightarrow$6$_{1,5}$   &  271529.02 &  35.5 & 0.050 (0.006)  & 0.041 (0.007) &  \\
$^{34}$SO$_2$   & 3$_{1,3}$$\rightarrow$2$_{0,2}$  & 102031.88 &  7.6  & 0.052 (0.003) &  &  \\
                & 5$_{1,5}$$\rightarrow$4$_{0,4}$  & 133471.43 &  15.5 & 0.041 (0.005) &  &  \\
                & 4$_{2,2}$$\rightarrow$4$_{1,3}$  & 141158.94 &  18.7 & 0.010 (0.004) &  &  \\            
                & 3$_{2,2}$$\rightarrow$2$_{1,1}$  & 203225.06 &  15.0 & 0.035 (0.010) &  &  \\
CCS                & 7$_6$$\rightarrow$6$_5$           & 86181.41  &  23.3 & 0.162 (0.003) & 0.111 (0.003) & \\
                & 7$_7$$\rightarrow$6$_6$           & 90686.38  &  26.1 & 0.133 (0.002) & 0.079 (0.002) & \\
                & 7$_8$$\rightarrow$6$_7$           & 93870.09  &  19.9 & 0.420 (0.002) & 0.238 (0.002) & \\
                & 8$_7$$\rightarrow$7$_6$           & 99866.50  &  28.1 & 0.119 (0.002) & 0.070 (0.002) & \\
                & 8$_8$$\rightarrow$7$_7$           & 103640.75 &  31.1 & 0.119 (0.002) & 0.059 (0.003) & \\
                & 8$_9$$\rightarrow$7$_8$           & 106347.73 &  25.0 & 0.286 (0.002) & 0.156 (0.002) & \\
                & 9$_8$$\rightarrow$8$_7$           & 113410.20 &  33.6 & 0.077 (0.002) & 0.047 (0.002) & \\
                & 9$_9$$\rightarrow$8$_8$           & 116594.78 &  36.7 & 0.090 (0.008) & 0.056 (0.010) & \\
                & 10$_{11}$$\rightarrow$9$_{10}$    & 131551.96 &  37.0 & 0.084 (0.004) & 0.066 (0.005) & \\
                & 11$_{10}$$\rightarrow$10$_9$      & 140180.74 &  46.4 & 0.032 (0.005) & 0.016 (0.005) & Red wing \\
                & 11$_{11}$$\rightarrow$10$_{10}$   & 142501.69 &  49.7 & 0.014 (0.004) & 0.013 (0.005) &  \\
                & 11$_{12}$$\rightarrow$10$_{11}$   & 144244.82 &  43.9 & 0.067 (0.004) &               &   \\
                & 12$_{13}$$\rightarrow$11$_{12}$   & 156981.65 &  51.5 & 0.032 (0.007) &               &   \\
CC$^{34}$S       &  7$_8$$\rightarrow$6$_7$ &  91913.53  & 19.5  & 0.015 (0.003)  & 0.011(0.003)  &    \\
                 &  8$_8$$\rightarrow$7$_7$ & 101371.04  & 30.6  & 0.005 (0.001)  &               &    \\
                 &  8$_9$$\rightarrow$7$_8$ & 104109.33  & 24.5  & 0.006 (0.002)  & 0.006 (0.002) &    \\                 
CCCS            & 15$\rightarrow$14  & 86708.38  &  33.3  & 0.052 (0.003) & 0.205 (0.003) & overlapped with HCO  \\
                & 16$\rightarrow$15  & 92488.49  &  37.7  & 0.034 (0.003) & 0.022 (0.003) &  \\
                & 17$\rightarrow$16  & 98268.52  &  42.4  & 0.021 (0.002) & 0.015 (0.002) &  \\
                & 18$\rightarrow$17  & 104048.45 &  47.4  & 0.017 (0.002) &               &   \\
                & 19$\rightarrow$18  & 109828.29 &  52.7  & 0.015 (0.003) &               &   \\
C$^{13}$CCS     & 15$\rightarrow$14  & 85838.10  &  33.0  & 0.006 (0.002) &               &    \\              
o-H$_2$S        &  1$_{1,0}$$\rightarrow$1$_{0,1}$  & 168762.75  &  8.1  & 0.892 (0.016) &  0.381 (0.019) & Red wing \\
NS          &  $^2\pi_{1/2}$ 5/2$_{1,7/2}$$\rightarrow$3/2$_{-1,5/2}$   & 115153.93  & 8.8 & 0.236 (0.007) &  0.393 (0.008) &     \\
            &  $^2\pi_{1/2}$ 5/2$_{1,5/2}$$\rightarrow$3/2$_{-1,3/2}$  & 115156.81  & 8.8 & 0.141 (0.007) &  0.273 (0.008) &     \\
            &  $^2\pi_{1/2}$ 5/2$_{1,3/2}$$\rightarrow$3/2$_{-1,1/2}$  & 115162.98  & 8.8 & 0.049 (0.011) &  0.166 (0.013) & Bad baseline \\
            &  $^2\pi_{1/2}$ 5/2$_{1,3/2}$$\rightarrow$3/2$_{-1,3/2}$  & 115185.34  & 8.8 & 0.040 (0.009) &                &  \\
            &  $^2\pi_{1/2}$ 5/2$_{1,5/2}$$\rightarrow$3/2$_{-1,5/2}$  & 115191.46  & 8.8 &               &  0.068 (0.012) &   \\
            &  $^2\pi_{1/2}$ 5/2$_{-1,5/2}$$\rightarrow$3/2$_{1,5/2}$  & 115489.41  & 8.9 & 0.067 (0.009) &                &   \\
            &  $^2\pi_{1/2}$ 5/2$_{-1,3/2}$$\rightarrow$3/2$_{1,3/2}$  & 115524.60  & 8.9 & 0.036 (0.009) &                &  \\
            &  $^2\pi_{1/2}$ 5/2$_{-1,7/2}$$\rightarrow$ 3/2$_{1,5/2}$ & 115556.25  & 8.9 & 0.471 (0.010) &                &  \\
            &  $^2\pi_{1/2}$ 5/2$_{-1,5/2}$$\rightarrow$3/2$_{1,3/2}$  & 115570.76  & 8.9 & 0.318 (0.009) &                & \\
            &  $^2\pi_{1/2}$ 5/2$_{-1,3/2}$$\rightarrow$3/2$_{1,1/2}$  & 115571.95  & 8.9 & 0.194 (0.008) &                &  \\
\hline \hline 
\end{longtable}

\begin{landscape}

\begin{figure}[h]
\hspace*{-0.5cm}
\includegraphics[scale=0.90,angle=0]{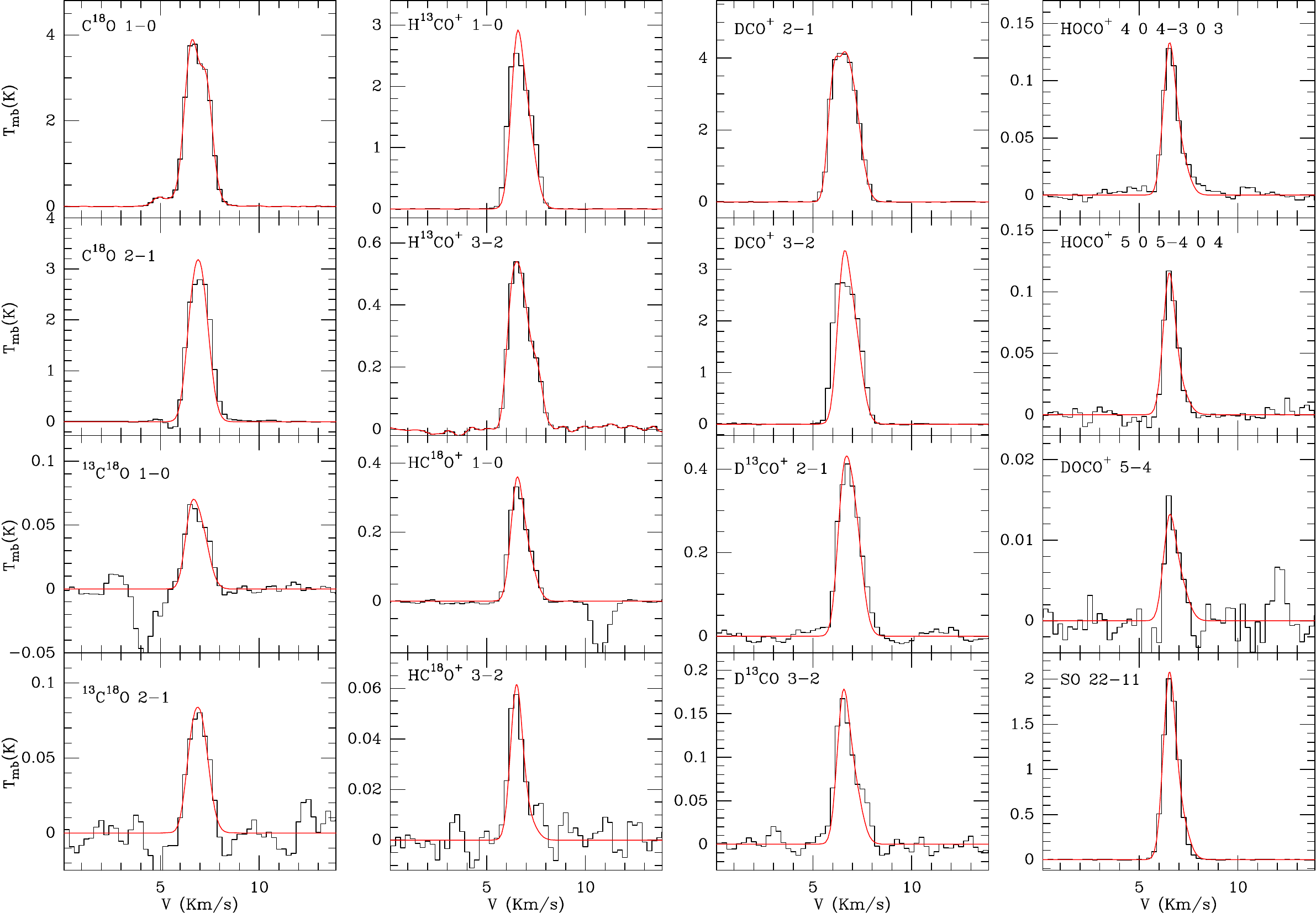}
   \caption{\label{figB1}
Spectra as observed with the IRAM 30-m telescope. The scale is in main beam brightness temperature. Negative features are artefacts 
of the frequency switching observing procedure. In red, the fits shown in 
Table B.1.
}
\end{figure}

\begin{figure}[h]
\hspace*{-0.5cm}
\includegraphics[scale=0.90,angle=0]{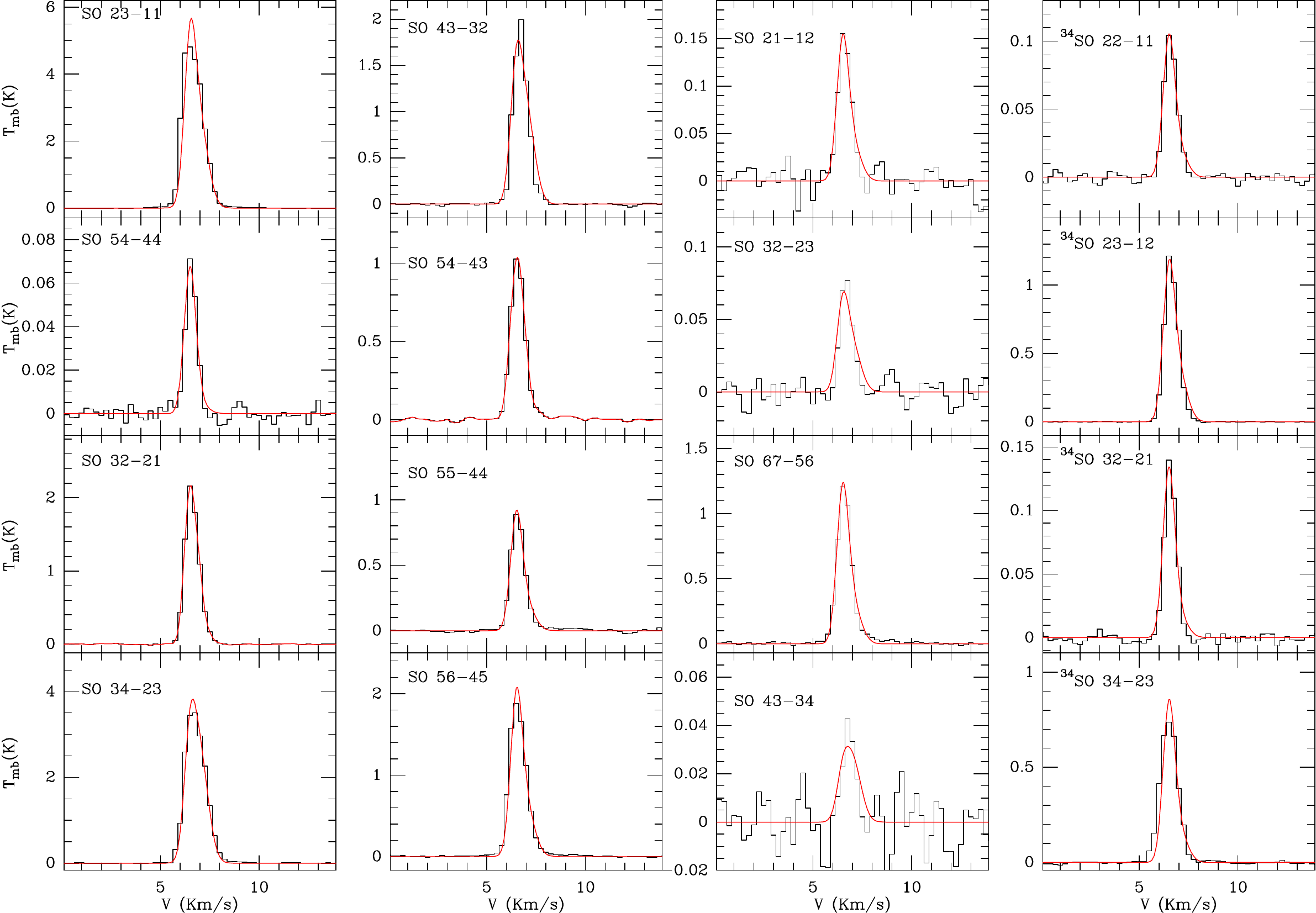}
   \caption{\label{FigB2}
The same as Fig~\ref{figB1}
}
\end{figure}

\begin{figure}[h]
\hspace*{-0.5cm}
\includegraphics[scale=0.90,angle=0]{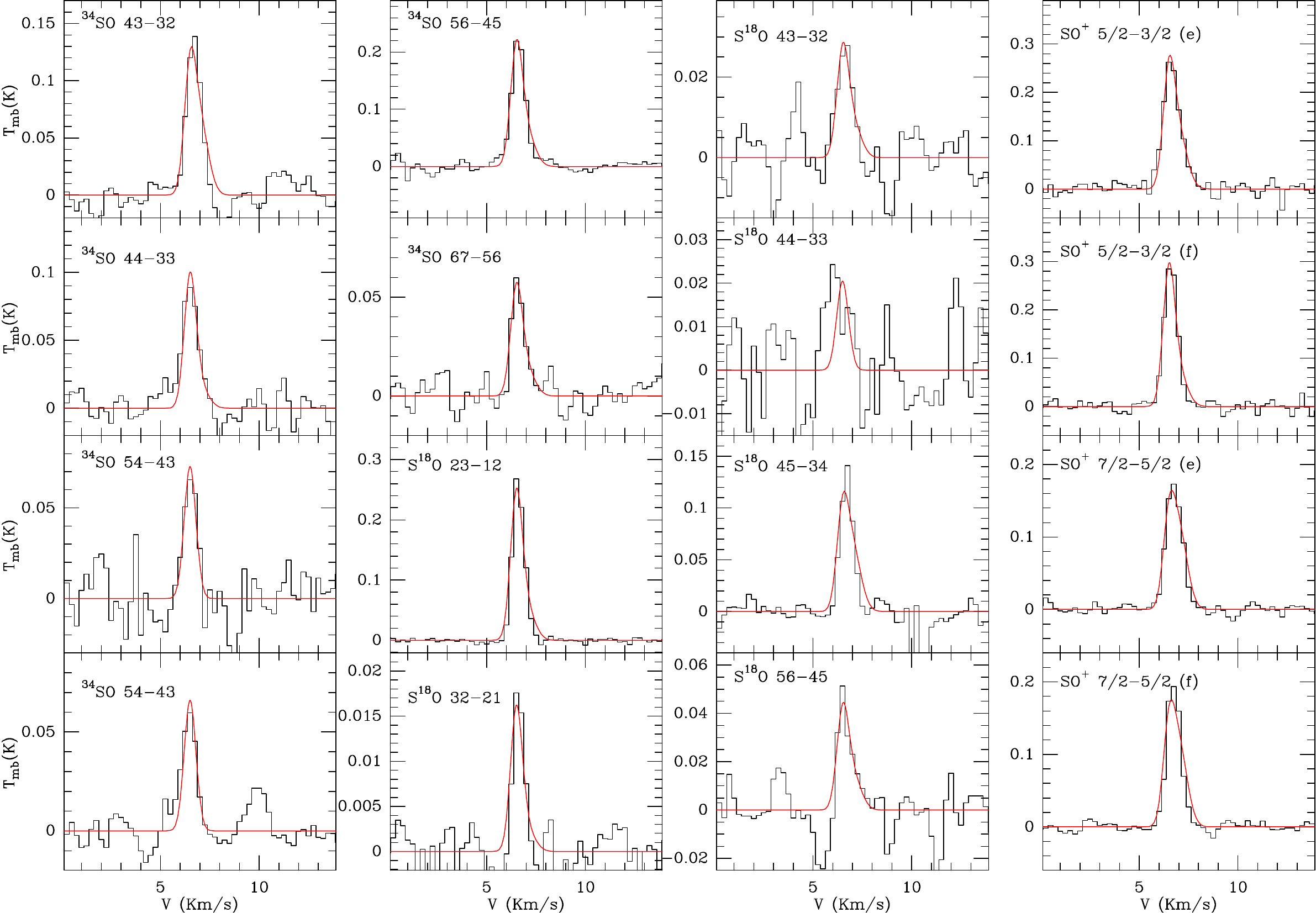}
   \caption{\label{FigB3}
The same as Fig~\ref{figB1}
}
\end{figure}

\begin{figure}[h]
\hspace*{-0.5cm}
\includegraphics[scale=0.90,angle=0]{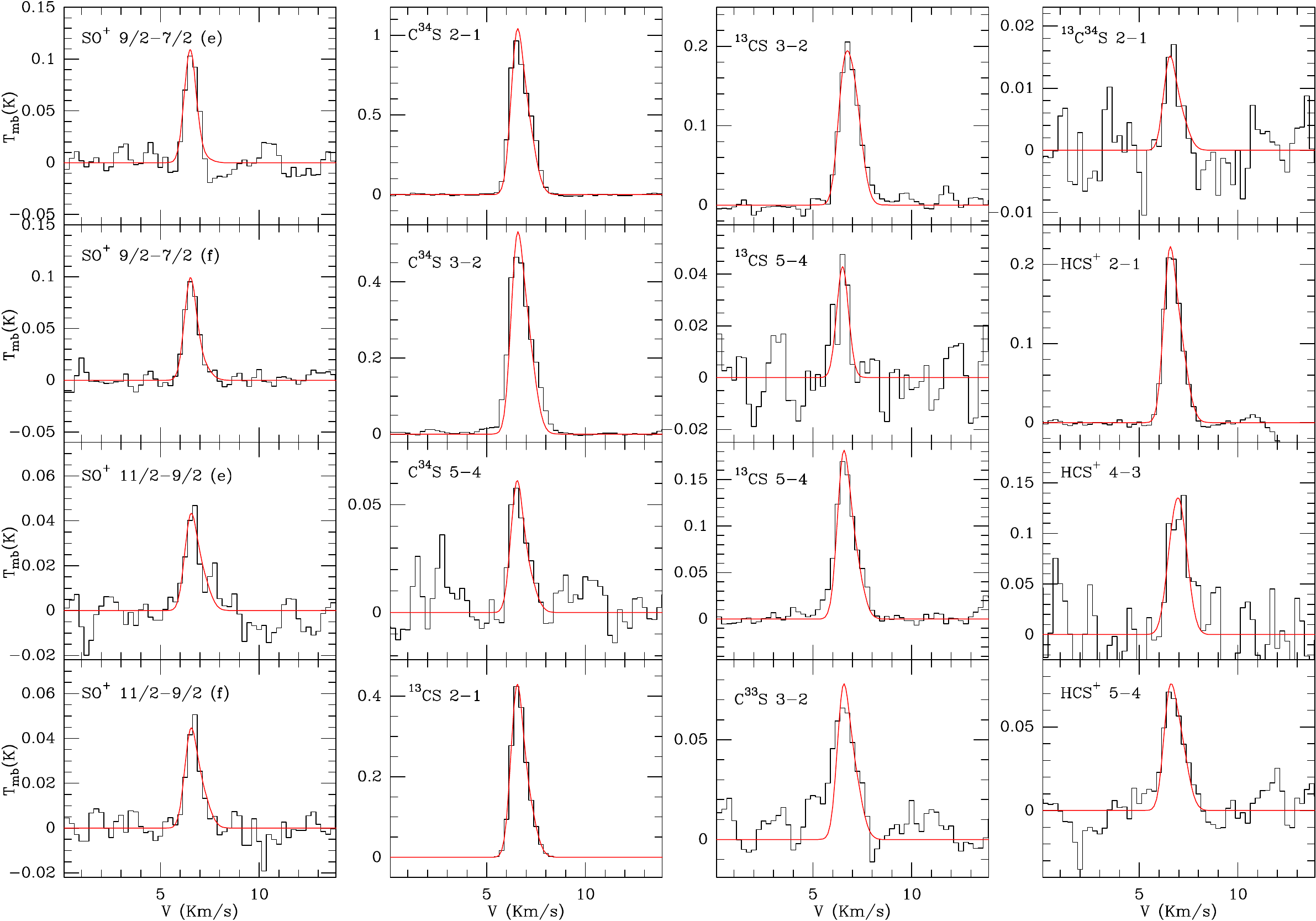}
   \caption{\label{FigB4}
The same as Fig~\ref{figB1}
}
\end{figure}

\begin{figure}[h]
\hspace*{-0.5cm}
\includegraphics[scale=0.90,angle=0]{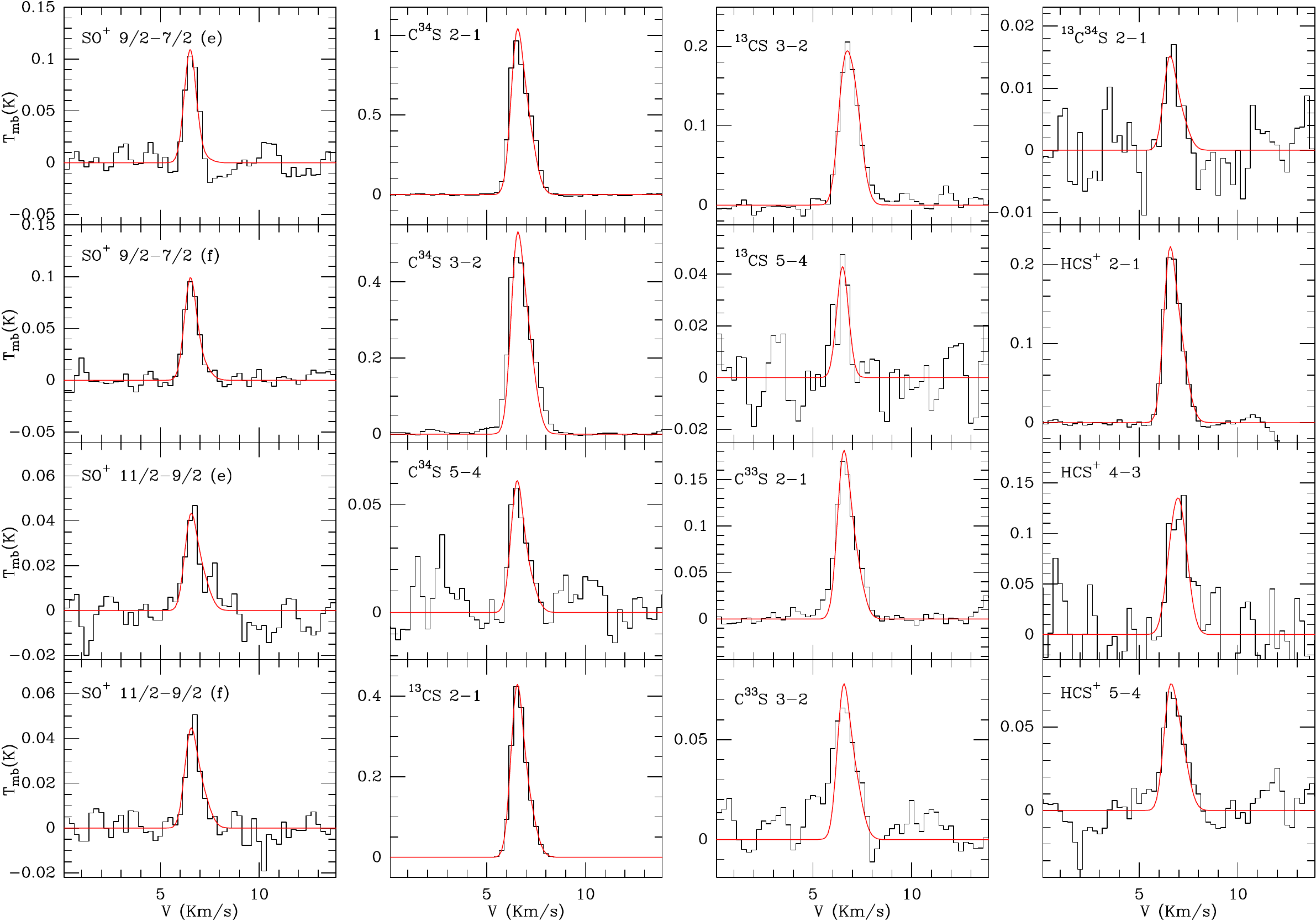}
   \caption{\label{FigB5}
The same as Fig~\ref{figB1}
}
\end{figure}

\begin{figure}[h]
\hspace*{-0.5cm}
\includegraphics[scale=0.90,angle=0]{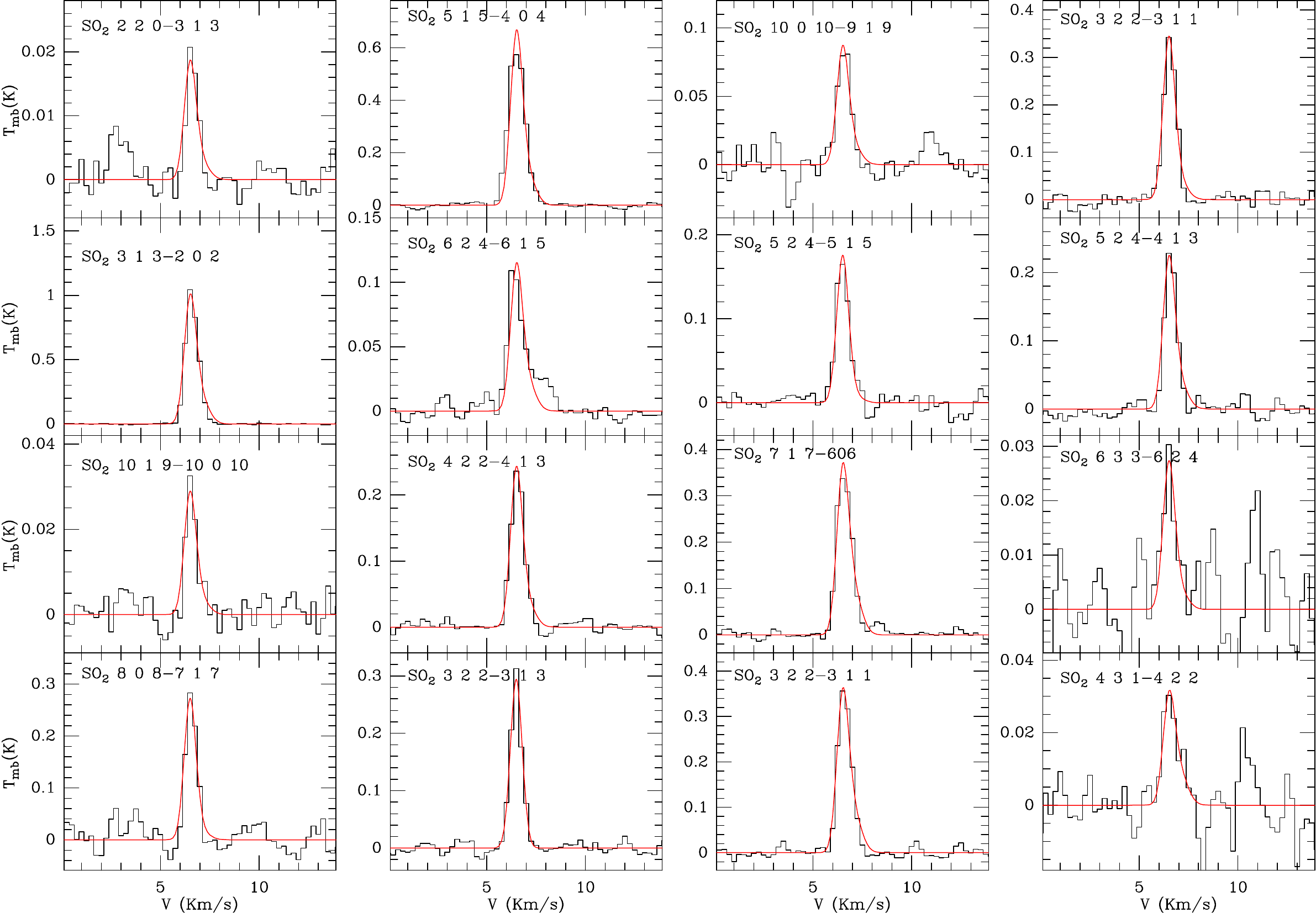}
   \caption{\label{FigB6}
The same as Fig~\ref{figB1}
}
\end{figure}

\begin{figure}[h]
\hspace*{-0.5cm}
\includegraphics[scale=0.90,angle=0]{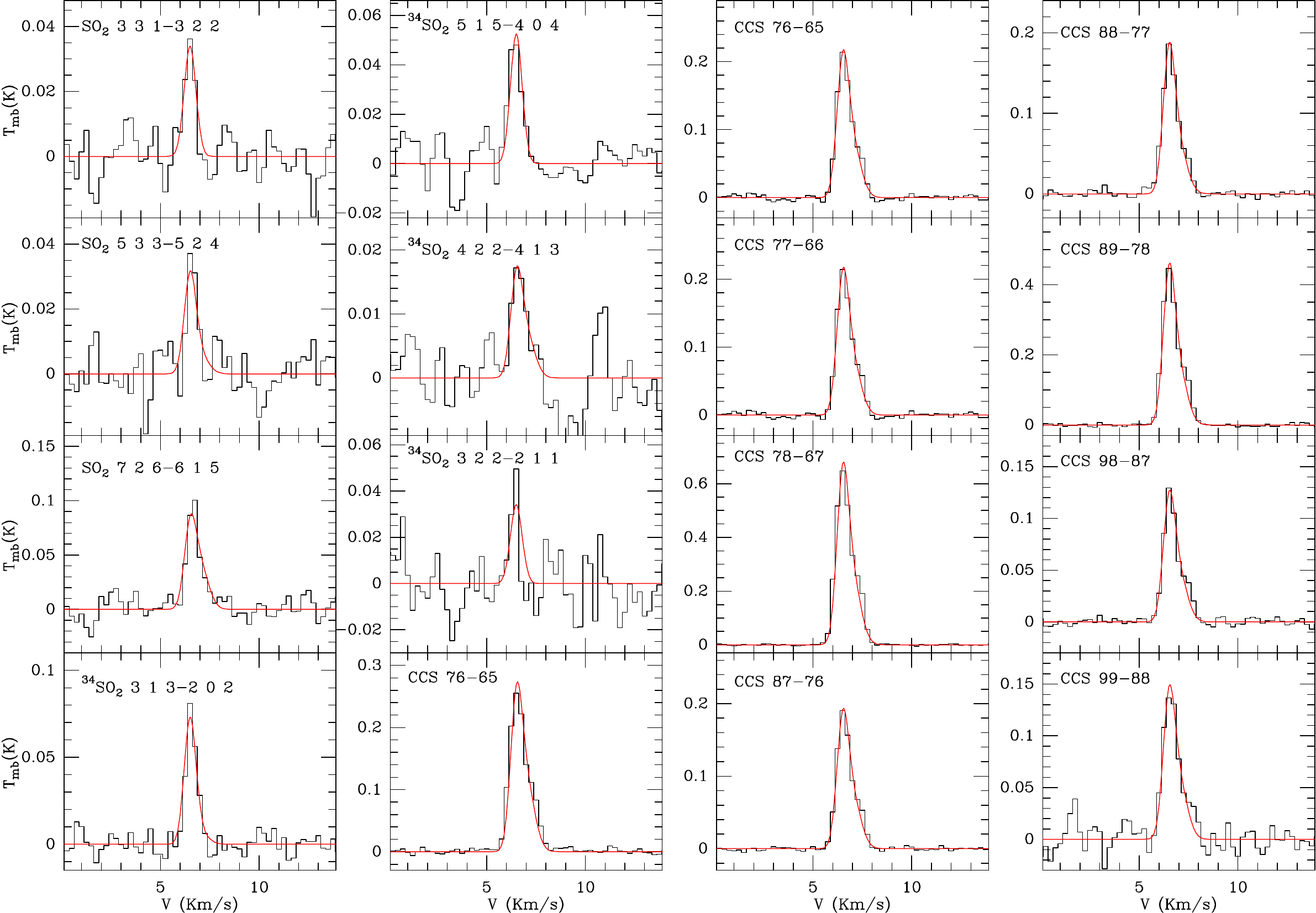}
   \caption{\label{FigB7}
The same as Fig~\ref{figB1}
}
\end{figure}

\begin{figure}[h]
\hspace*{-0.5cm}
\includegraphics[scale=0.90,angle=0]{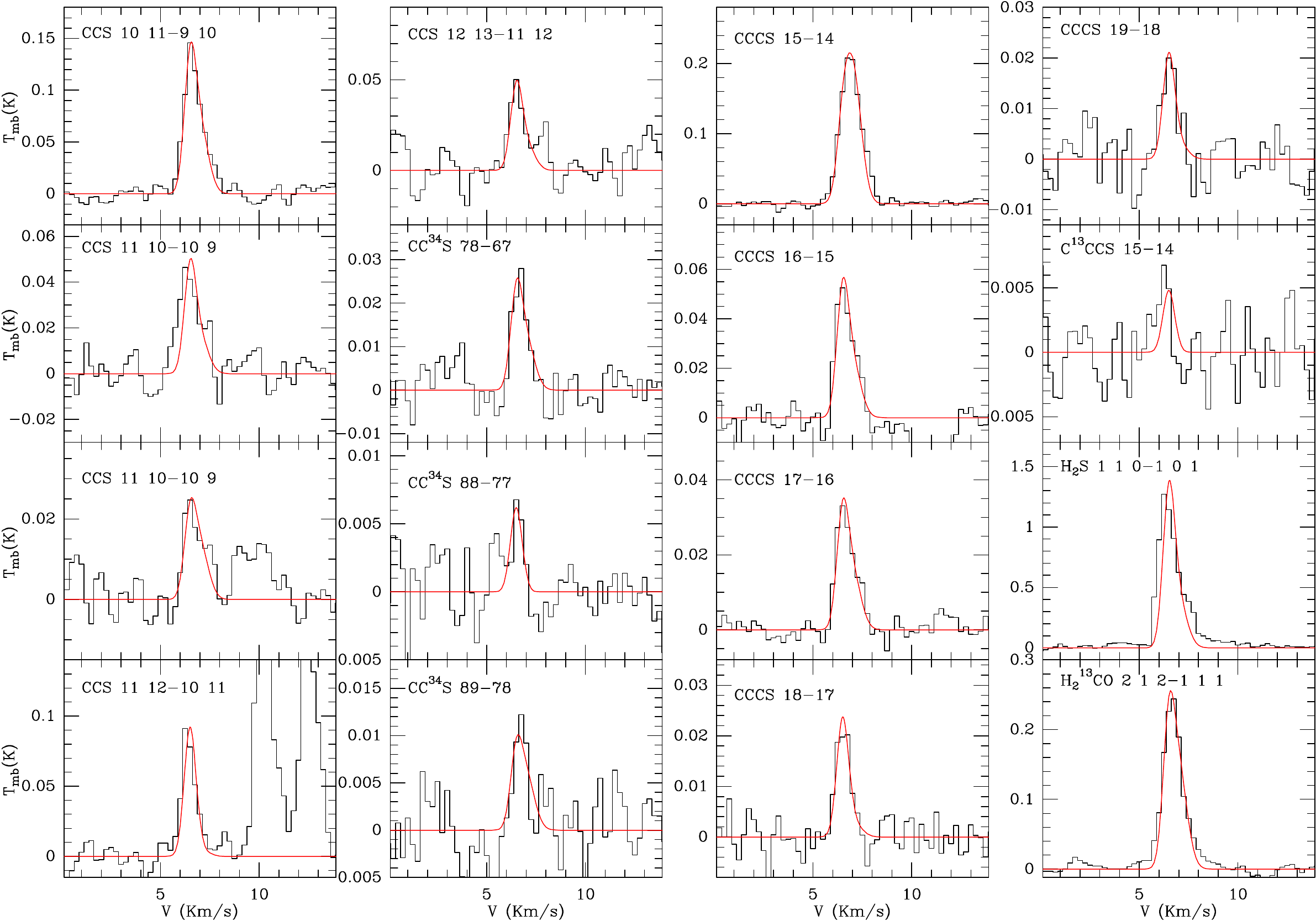}
   \caption{\label{FigB8}
The same as Fig~\ref{figB1}
}
\end{figure}

\begin{figure}[h]
\hspace*{-0.5cm}
\includegraphics[scale=0.90,angle=0]{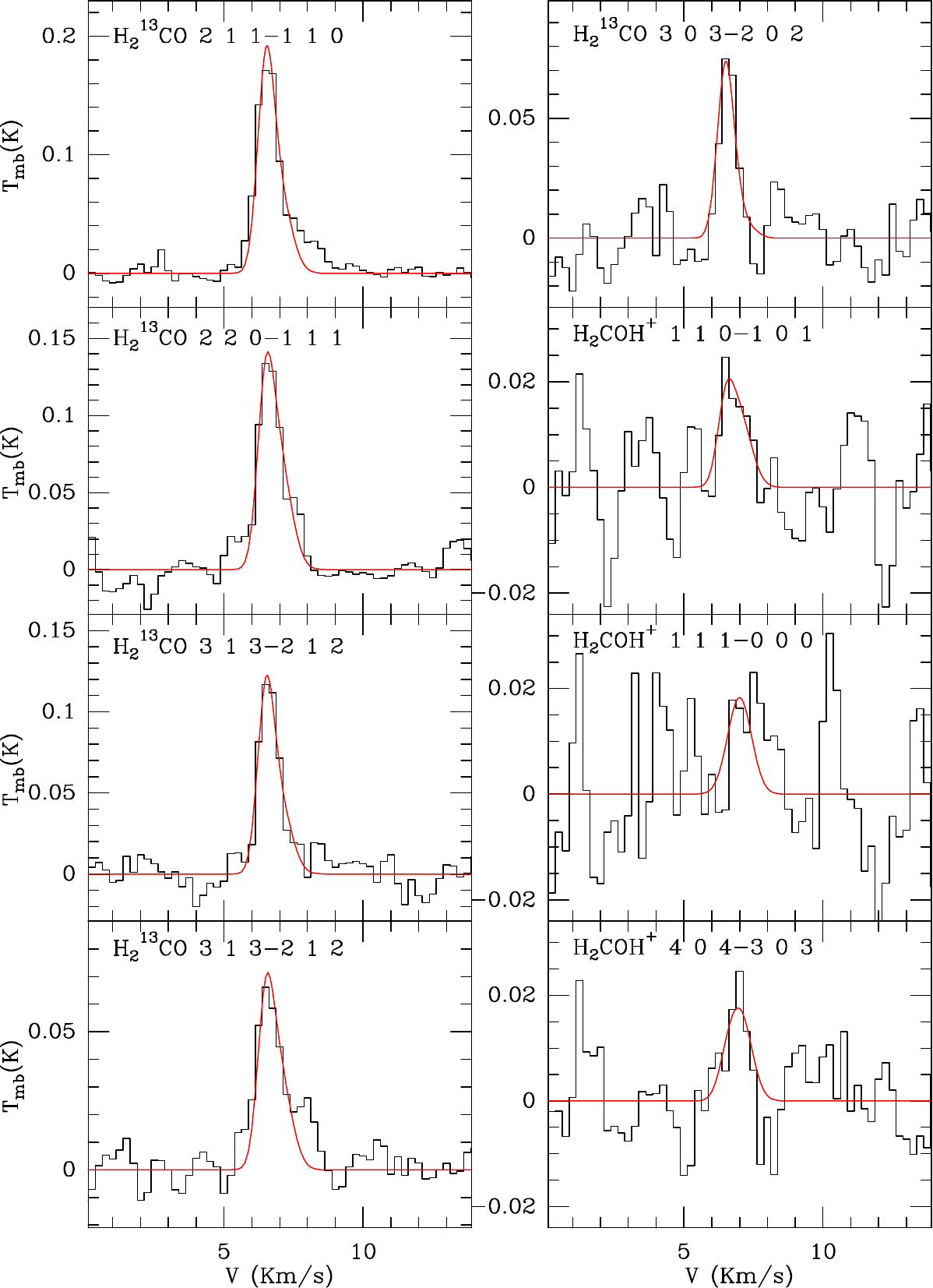}
   \caption{\label{figB9}
The same as Fig~\ref{figB1}
}
\end{figure}

\end{landscape}

\end{document}